\documentclass[aps,twocolumn,superscriptaddress,showpacs,showkeys]{revtex4-2}
\DeclareUnicodeCharacter{2212}{\ensuremath{-}}
\usepackage[T1]{fontenc}
\usepackage{textcomp}

\usepackage{graphicx}
\usepackage{bibunits}
\usepackage[utf8x]{inputenc}
\usepackage{hyperref}
\usepackage{xcolor}
\usepackage{etoolbox}
\usepackage{xcolor}
\usepackage{amsmath}
\usepackage{upgreek}
\usepackage{soul}
\usepackage{float}
\usepackage{placeins}
\usepackage{afterpage}
\hypersetup{
    colorlinks,
    linkcolor={blue}, 
    citecolor={blue},
    urlcolor={blue!80!black}
}

\begin{document}

\title{Carrier localization and dynamics in  In$_{0.10}$Ga$_{0.90}$N: the impact of alloying and Si doping}

\author{O. Mpatani}
\email{ongeziwe.mpatani@cern.ch}
\affiliation{Materials Physics Research Institute, School of Physics, University of the Witwatersrand, Johannesburg, Wits 2050, South Africa}
 
\author{D. Muth} \author{A. Krüger}
\affiliation{Department of Physics and Marburg Center for Quantum Materials and Sustainable Technologies, Semiconductor Spectroscopy Group, Philipps-Universit\"{a}t Marburg, 35032 Marburg, Germany}

\author{R. Adhikari}
\affiliation{Institute of Semiconductor and Solid-State Physics, Johannes Kepler University, A-4040 Linz, Austria}
\affiliation{Linz Institute of Technology, Johannes Kepler University, A-4040 Linz, Austria}

 \author{A. Bonanni}
\affiliation{Institute of Semiconductor and Solid-State Physics, Johannes Kepler University, A-4040 Linz, Austria}

\author{M. Gerhard}
\affiliation{Department of Physics and Marburg Center for Quantum Materials and Sustainable Technologies, Semiconductor Spectroscopy Group, Philipps-Universit\"{a}t Marburg, 35032 Marburg, Germany}

\author{H. Masenda}
\email{hilary.masenda@wits.ac.za}
\affiliation{Materials Physics Research Institute, School of Physics, University of the Witwatersrand, Johannesburg, Wits 2050, South Africa}

\date{\today}

\begin{abstract}

Alloying and doping are crucial for enhancing the electronic and optical properties of semiconductors while simultaneously introducing disorder. This report explores the effects of alloying and Si (0.5 at.\%) doping on In$_{0.10}$Ga$_{0.90}$N thin films that were grown by metal-organic vapor phase epitaxy. Post-growth X-ray diffraction measurements indicate that Si doping does not affect the lattice parameters and screw dislocations but significantly increases the edge dislocation density. Temperature-dependent time-resolved photoluminescence spectroscopy shows that Si-doped In$_{0.10}$Ga$_{0.90}$N exhibits higher photoluminescence intensity, blue-shifted peaks, narrower emission linewidths, and quenching of lower energy sidebands when compared to pristine In$_{0.10}$Ga$_{0.90}$N. The peak energies of the most dominant feature, the donor-bound exciton, for both samples show an $S$-shape behavior indicating the presence of disorder. Although doping improves luminescence, it also introduces deeper localized states. This suggests that impurity-induced disorder outweighs compositional fluctuations, as confirmed by higher disorder parameters and Stokes shifts. Thus, the Si doping leads to increased localization, reducing nonradiative recombination channels while enhancing radiative processes. The deeper states in the doped sample confirm improved carrier confinement, and their saturation leads to early thermalization, thereby lowering the red-blue shift transition from 165 K to about 50 K. Even though the high doping level makes Si-doped In$_{0.10}$Ga$_{0.90}$N a degenerate system, it exhibits enhanced luminescence properties. These findings shed light on the impact of silicon doping on charge transport in InGaN alloys for optoelectronic applications.
\end{abstract}

\maketitle

\section{\label{sec:level1}Introduction}

Alloys of semiconductors are essential materials for tuning properties in heterostructures used in electronic and optical devices. {In$_{x}$Ga$_{1-x}$N} is an alloy of GaN and InN that is achieved by controlling the relative compositions of indium and gallium. Remarkable progress has been made with the ability to adjust the band gap ranging from 0.7 eV in InN to 3.4 eV in GaN\cite{Nakamura2000,Morkoc2009,Adachi2009,moustakas2017optoelectronic}. The ternary alloy is widely used in light-emitting diodes (LED) and laser diodes (LD)\cite{nakamura2013history,chen2020recent,hardy2011group}. The 2014 Nobel Prize in Physics recognized the technological significance of efficient blue LEDs\cite{nakamura2015nobel,akasaki2015nobel,Amano2015nobel}, which rely heavily on the InGaN active layers as the essential component\cite{cai2023red, funato2013remarkably}.

Moreover, compositionally graded ternary nitrides offer a practical approach to creating three-dimensional electron or hole slabs by utilizing polarization-induced doping through elemental compositional gradients\cite{Rajan2006,Brown2010,Adhikari2016,Sang2021,zhang2021compositionally}. This approach can be used to tailor the properties of III-nitride layers in heterostructures for devices such as high-electron-mobility transistors (HEMTs)\cite{grandpierron2024understanding} or photovoltaic cells\cite{Brown2010,zhao2023toward}. By controlling the spatial distribution of group III elemental concentrations during growth, the properties related to carrier transport can be significantly improved.

Alloying has proven to be beneficial in altering the structural properties of semiconducting materials, which in turn allows for the tuning of their electronic and optical properties\cite{Adachi2009}. However, this process comes at the cost of introducing compositional disorder\cite{baranovski2006charge,baranovskii2022energy,David-Weisbusch2022}. This disorder arises from the stochastic distribution of atoms within the lattice, leading to areas where specific elements are spatially concentrated. This introduces localized states\cite{Anderson1958} that create a potential barrier which affects carrier transport when alloys are used in devices\cite{nakamura1998,O'Donnell1999,Weisbuch2021,wu2021Anderson,Weisbuch2021,Bhunia2024,Xue2025}. The theory behind the localization landscape of quantum disorder effects in semiconductors has recently been extensively explored\cite{filoche2017localization,piccardo2017localization,li2017localization,gebhard2023quantum,nenashev2023quantum}, covering modeling and applications in quantum well (QW) layers, along with their influence on carrier transport and recombination in LEDs.

Furthermore, the intentional addition of impurities or foreign atoms plays a vital role in altering semiconducting materials and tailoring their properties for specific applications. The bonding mechanism between valence electrons, the resulting atomic arrangements, and any associated defects influence the properties of modified semiconducting materials. Doping is used to induce both p-type and n-type conductivity, as well as to alter resistivity\cite{Adachi2009}. In III-nitrides (and metal oxides), traditionally, it has been a challenge to p-dope; in GaN, this has been achieved with Mg\cite{Amano1989,Nakamura1992Hole}. Additionally, over the past two decades, interest in nitrides grew following the prediction of room-temperature ferromagnetism when doped with transition metals and achieving p-doping with the right concentration of holes\cite{dietl2000zener,Dietl2014,Dietl2019,Bonanni2021}. Silicon is recognized as the most effective and widely used n-type dopant in III-nitrides, attributed to the high carrier concentrations achievable\cite{pampili2017doping}. Moreover, silicon doping has been explored as a way to enhance the optical and structural properties of InGaN. Studies on Si-doped InGaN/GaN nanowires have reported improved morphology, reduced defect densities, enhanced thermal stability, and better interface quality\cite{bose2018imaging}. Optically, Si-doping in QWs results in blue-shifted photoluminescence (PL) peaks, reduced Stokes shifts, shorter PL decay times, and increased PL intensity. These improvements are linked to changes in band structure and recombination dynamics induced by silicon\cite{ryu2000silicon,cheng2003mechanisms}. In particular, Si-doping in QWs screens the internal piezoelectric field, thereby mitigating the quantum-confined Stark effect (QCSE), and enhancing radiative recombination rates. In addition to that, it may influence carrier localization depending on doping level and material quality\cite{chen2005effects,ryu2001effects,davies2016}. There is comparatively limited literature on the effects of silicon doping in InGaN films. Moon et al.\cite{moon1999optical} reported suppression of compositional fluctuations, which are typically present in InGaN alloys due to indium inhomogeneity. The authors also observed that indium-rich regions in the InGaN lattice expanded with increasing film thickness. Li et al.\cite{li2009suppression} found that silicon incorporation in {In$_{x}$Ga$_{1-x}$N} ($x$ = 32\%) led to phase separation within the films. In a separate study, Yamamoto et al.\cite{yamamoto2013} observed a significant reduction in indium concentration with increased silicon incorporation, which led to a wider band gap in InGaN as the silicon flow rate increased. This indicates that Si tends to limit In occupancy of the III sub-lattice sites rather than Ga sites when it is introduced during growth. One possible explanation for this preference is that silicon has a smaller atomic radius compared to indium. More importantly, Si-N bonds have lower formation energies than In-N bonds, which may contribute to the reduction of lattice strain\cite{yamamoto2013,pandey2013band}.

Although InGaN is widely used in commercial devices, its luminescence efficiency is influenced by high threading dislocation densities and compositional fluctuations\cite{dang2018threading,park2024ingan,kang2025growth}. Numerous studies have investigated the origins and effects of these factors on device performance\cite{ chichibu2006origin,ponce2003microstructure,haller2018gan}. PL is a non-destructive technique commonly used to probe optical properties such as band structure, point defects, and luminescence impurities. Information on nonradiative defects, localized states, and carrier transport dynamics can be obtained from PL studies as a function of temperature. Additionally, time-resolved PL (TRPL) enables the measurement of carrier lifetimes and helps distinguish between free, bound, and localized excitons, as well as tunneling effects\cite{lu2014temperature,pecharroman2004investigation,li2016time}.

This study investigates the effects of alloying and Si doping on optical properties in as-grown InGaN thin films. Temperature-dependent time-resolved PL (TD-TRPL) measurements were undertaken to explore the influence of alloying and Si doping on carrier localization and recombination dynamics. A lower concentration has been selected for this study because higher indium content ($x \approx 25 - 40\%$)\cite{prabakaran2022} can lead to structural degradation, phase separation, and a tendency to transition towards polycrystalline or In-rich nano-island phases. The allowable indium concentration limits for achieving crystalline In$_{x}$Ga$_{1-x}$N thin films differ based on the particular growth technique and specific conditions employed. 

\section{Experimental Details}
\label{sec:experimental details}

\subsection{MOVPE}

Pristine and silicon-doped (Si 0.5 at.\%) In$_{x}$Ga$_{1-x}$N thin films were synthesized using an AIXTRON 200RF horizontal tube metal-organic vapor phase epitaxy (MOVPE) reactor on $c$-plane sapphire substrates. These are labeled as In$_{0.10}$Ga$_{0.90}$N and In$_{0.10}$Ga$_{0.90}$N:Si, respectively. The precursors employed included trimethylgallium (TMGa), trimethylindium (TMIn), ammonia (NH$_3$), and silane (SiH$_4$), which act as sources for Ga, In, N, and Si, respectively. H$_2$ served as the carrier gas. Initially, the substrate is subjected to ammonia treatment for nitridation, followed by the growth of a GaN nucleation layer at 540$^{\circ}$C by introducing TMGa into the reactor, which was subsequently annealed at 1040$^{\circ}$C. Next, a 1 $\upmu$ m thick GaN buffer layer was grown at 1035$^{\circ}$C, at an operating pressure of 200 mbar. Finally, a 130 nm layer of In$_{x}$Ga$_{1-x}$N (or In$_{x}$Ga$_{1-x}$N:Si) was grown at 750$^{\circ}$C through the introduction of TMIn (with SiH$_4$ added for the doped sample) and reactor pressure of 300 mbar. To have real-time control over the entire fabrication process, the MOVPE system is equipped with an \textit{in situ} reflectometer that allows for both spectroscopic and kinetic measurements in the energy range $\sim$1.5--5.5 eV. The structures are characterized by high-resolution X-ray diffraction (HR-XRD) to obtain information on the surface crystal structure and dislocations.

\subsection{HR-XRD}

X-ray radial curves (XRC) and reciprocal space maps (RSM) were measured using a PANalytical X’Pert Pro Materials Research Diffractometer (MRD), equipped with a hybrid monochromator featuring a 1/4$^{\circ}$ divergence slit. This diffractometer is also fitted with a 256-channel solid-state PixCel detector with a 9.1 mm anti-scatter slit. The radial curves from the 2$\theta-\omega$ scan in Fig. \ref{fig:XRD}, were measured on the symmetric $(0002)$ and asymmetric $(20\bar{1}4)$ Bragg reflections of the wurtzite GaN crystal system. These were utilized to establish the overall layer structure and identify the crystallographic phases. 

\subsection{Photoluminescence}

TD-TRPL measurements were performed on as-grown In$_{0.10}$Ga$_{0.90}$N and In$_{0.10}$Ga$_{0.90}$N:Si films to investigate their optical properties over a temperature range of 10 to 300 K. An 80 MHz titanium-sapphire laser (Spectra Physics, Tsunami) was used throughout. It is an excitation source with wavelengths in the range 680 - 1040 nm with a pulse duration of $\sim$100 fs. A second-harmonic generator (CSK Optronics, SuperTripler) allowed for a wider wavelength range, extending it to the blue and ultraviolet regions. For these measurements, a central wavelength of 380 nm at a power level of 2 mW was used. The samples were mounted on a cold finger to ensure precise temperature regulation within a liquid-helium-flow microscopy cryostat (CryoVaC). The PL signal was collected in reflection geometry and detected using a streak camera (Hamamatsu) for spectral and temporally resolved spectra. Throughout the experiment, the detection system maintained a spectral resolution of 0.4 nm and a temporal resolution of 5 ps. This thorough approach enabled an in-depth study of the optical properties of In$_{0.10}$Ga$_{0.90}$N and In$_{0.10}$Ga$_{0.90}$N:Si across varying temperatures.

\section{Results and Discussion}
\label{sec:Results and Discussion}

\subsection{Lattice parameters and dislocation analysis}

\begin{figure}[h]
\includegraphics[width=0.49\textwidth]{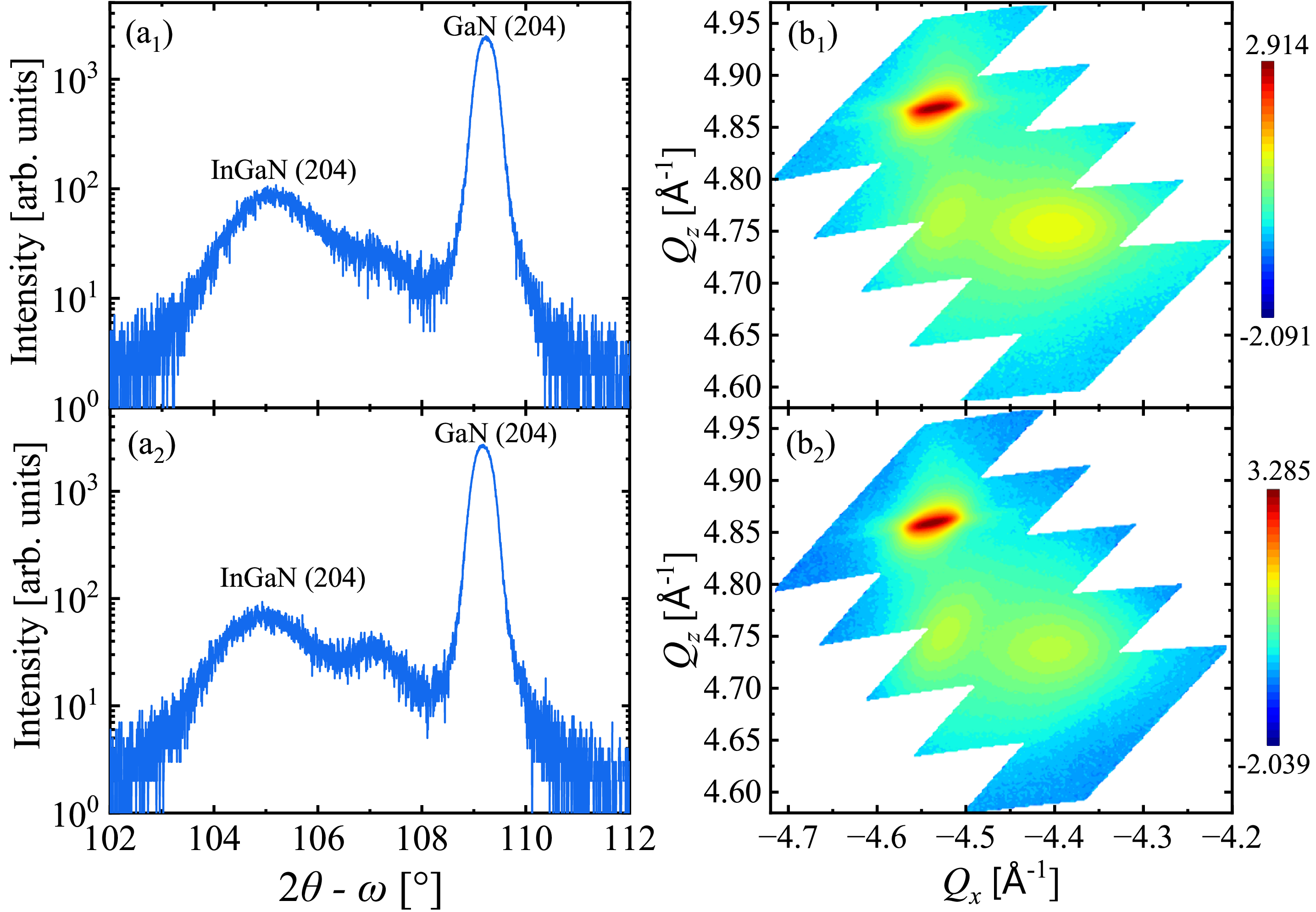} 
\caption{HR-XRD: (a) 2$\theta-\omega$ scan  and (b) reciprocal space maps of In$_{0.10}$Ga$_{0.90}$N (top) and In$_{0.10}$Ga$_{0.90}$N:Si (bottom).}
\label{fig:XRD}
\end{figure} 

The 2$\theta-\omega$ scans shown in Fig. \ref{fig:XRD}(a$_1$) and (a$_2$) correspond to In$_{0.10}$Ga$_{0.90}$N and In$_{0.10}$Ga$_{0.90}$N:Si, respectively. As expected for a layered thin film, there is an intense peak from the diffraction of X-rays on the buffer layer, GaN, situated at 109$^{\circ}$. Additionally, a less prominent feature at 105$^{\circ}$ can be observed for In$_{0.10}$Ga$_{0.90}$N or In$_{0.10}$Ga$_{0.90}$N:Si. These patterns confirm the successful synthesis of In$_{0.10}$Ga$_{0.90}$N and In$_{0.10}$Ga$_{0.90}$N:Si on the GaN buffer layer, indicating that high-quality thin films with good periodicity have been obtained. Furthermore, both XRD patterns exhibit a fringe at 107$^{\circ}$, which is more visible for Fig. \ref{fig:XRD}(a$_2$) than in \ref{fig:XRD}(a$_1$). Such fringes are often observed when high-resolution optics are utilized in thin films, and their presence suggests good crystal quality\cite{harrington2021back}. The XRD peaks allowed for the determination of the lattice parameters $a$ and $c$. Additionally, RSM was employed to monitor the evolution of these lattice parameters as well as dislocation density. All extracted values are summarized in Table \ref{tab:XRD}. 

The in-plane lattice parameters are nearly identical between the pristine and Si-doped samples, both averaging 3.195 $\text{\AA}$. The rocking curve and RMS measurements show excellent agreement within 0.01 to 0.03 $\text{\AA}$. This confirms Si doping does not significantly affect the in-plane lattice parameter. The $c$-lattice parameter, measured using multiple reflections, has average values of 5.180 $\text{\AA}$ for InGaN and 5.183 $\text{\AA}$ for InGaN:Si. The out-of-plane lattice parameters across different reflections are consistent and indicate that Si doping causes a negligible change of about 0.003 $\text{\AA}$. The similar lattice parameters demonstrate that both materials experience comparable strain, consistent with 10\% indium incorporation. Reported theoretical values for wurzite GaN are \textit{a} = 3.1878 $\text{\AA}$ and \textit{c} = 5.1850 $\text{\AA}$ \cite{Moram2009,Roccaforte2020}, confirming lattice expansion. This suggests that adding Si does not substantially impact In incorporation efficiency.

\begin{table*}[t]
    \caption{X-ray rocking curve, reciprocal space map and dislocation analysis parameters}
    \begin{tabular}{p{2.5cm} p{1.8cm} p{2cm} p{1.8cm} p{1.8cm} p{2cm} p{2.2cm} p{2.2cm}}
    \hline
        Sample & $a$ [204] ($\text{\AA}$) & $a$ [RSM] ($\text{\AA}$) & $c$ [002] ($\text{\AA}$) & $c$ [004] ($\text{\AA}$) & $c$ [RSM] ($\text{\AA}$) & Screw (cm$^{-2}$) & Edge (cm$^{-2}$) \\
         \hline
        In$_{0.10}$Ga$_{0.90}$N & 3.18 & 3.21 & 5.19 & 5.19 & 5.16 & 5.22$\times10^{8}$ &2.23$\times10^{9}$ \\
         \hline
        In$_{0.10}$Ga$_{0.90}$N:Si & 3.19 & 3.20 & 5.18 & 5.19 & 5.18 & 5.21$\times10^{8}$ & 2.76$\times10^{9}$ \\
         \hline
    \end{tabular}
    \label{tab:XRD}
\end{table*}

In semiconducting materials, one of the primary defect types is dislocations, edge and screw, both of which create localized disruptions within the crystal structure. A crucial aspect of understanding these dislocations is the Burgers vector, which defines both the magnitude and direction of the lattice distortion that occurs as a result of the dislocation in the crystal lattice. The dislocation density gives an indication of the quality of the material. The screw dislocation densities are identical for both In$_{0.10}$Ga$_{0.90}$N and In$_{0.10}$Ga$_{0.90}$N:Si. The unchanged density suggests that Si doping does not introduce additional screw dislocations. Screw dislocations have Burgers vectors parallel to the $c$-axis and are likely to have an impact on electrical properties and vertical carrier transport. However, the edge dislocation density increases with Si by approximately 24\% (0.53$\times10^{9}$ cm$^{-2}$). The Burgers vectors are in the basal plane for edge dislocations. Thus, the incorporation of Si introduces additional structural changes. The combined dislocation densities are 2.75$\times10^{9}$ cm$^{-2}$ and 3.28$\times10^{9}$ cm$^{-2}$ for the pristine and doped samples, respectively, resulting in a total dislocation density increase of $\approx$19\%. Overall, dislocation densities in the order of $10^{9}$ cm$^{-2}$, confirm the good quality samples considering there is 10\% indium concentration\cite{ozturk2014microstructural}. Higher dislocation densities are in In$_{0.10}$Ga$_{0.90}$N when compared to pure GaN due to lattice mismatch between InN and GaN, phase separation tendencies, or compositional fluctuations. Thus, Si doping has minimal effect on lattice parameters and screw dislocation densities but significantly increases edge dislocation densities while maintaining the crystal quality of the prepared samples.

\subsection{Luminescence: effect of alloying and doping}

\begin{figure}[h]
\includegraphics[width=0.45\textwidth]{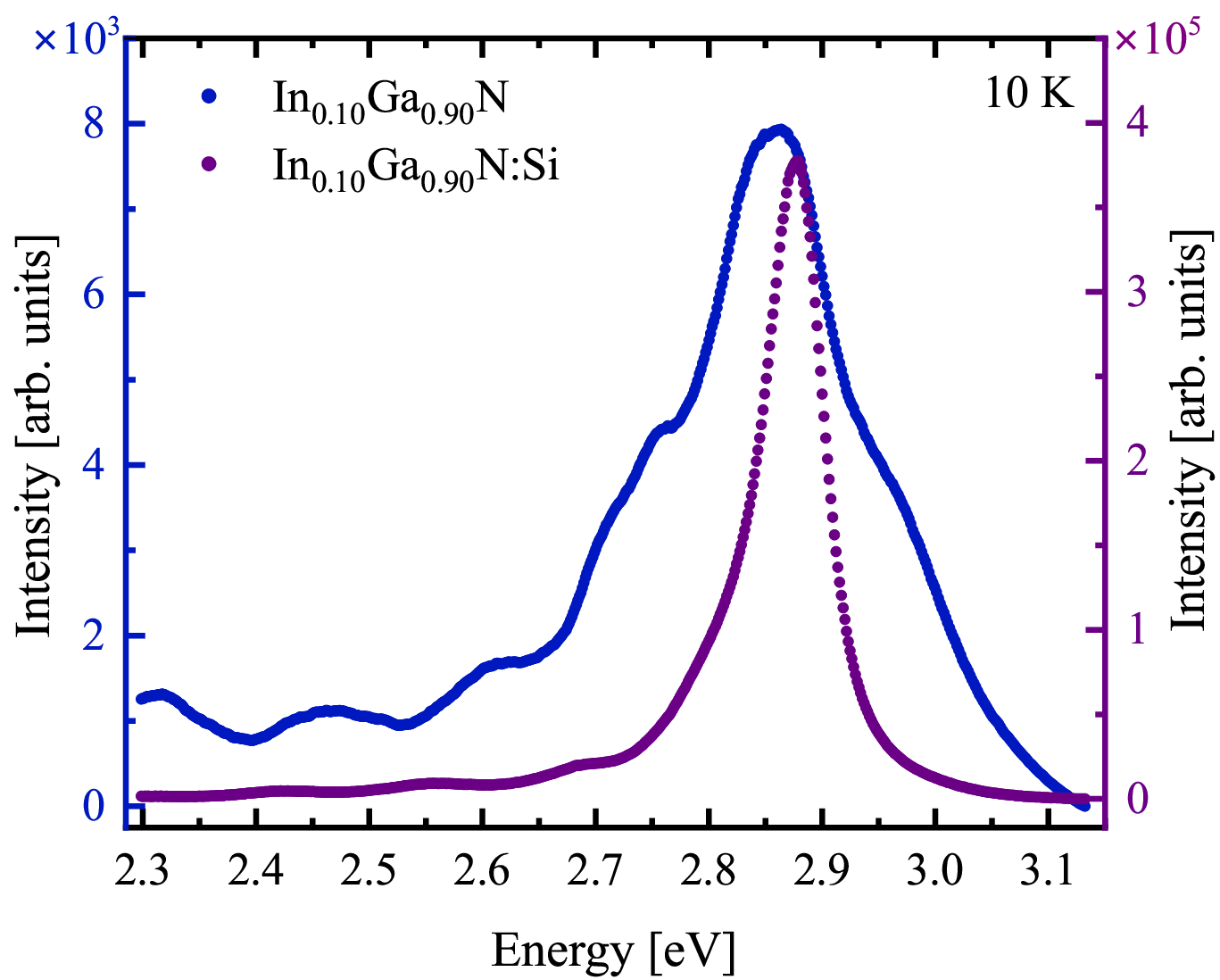}
\caption{\label{fig:InGaN}{The PL spectra of In$_{0.10}$Ga$_{0.90}$N and In$_{0.10}$Ga$_{0.90}$N:Si measured 10 K.}}
\end{figure}

The PL spectra of In$_{0.10}$Ga$_{0.90}$N and In$_{0.10}$Ga$_{0.90}$N:Si measured at 10 K are shown in Fig. \ref{fig:InGaN}. Both spectra exhibit asymmetric line shapes with additional shoulders (``bumps'') on either side of the more pronounced peak. The main differences observed between the two spectra are a blueshifted PL line, increased intensity, narrower full width at half maximum (FWHM) or linewidth, and diminished lower-energy sidebands resulting from Si incorporation.  The dominant emission peak for In$_{0.10}$Ga$_{0.90}$N is centered at 2.86 eV, with a linewidth of 95.62 meV, while that of In$_{0.10}$Ga$_{0.90}$N:Si is at 2.88 eV and has a narrower linewidth of 48.81 meV. Similar trends have been reported in the literature. Nakamura et al.\cite{nakamura1993si} studied Si-doped In$_{0.14}$Ga$_{0.86}$N at room temperature and observed sharp band-edge emission, which they attributed to shallow donor levels introduced by silicon. These donor levels likely promote enhanced radiative recombination, leading to increased PL intensity. Li et al.\cite{li2006influence,li2006enhanced} and Lu et al.\cite{lu2013influence} found that increasing Si content in InGaN progressively increased PL intensity while reducing the linewidth. They proposed two possible mechanisms for this behavior: (1) improved overlap between electron and hole wavefunctions due to increased carrier density, and (2) improved crystal quality. In a way, both mechanisms could be contributing to the observed effects.

\begin{figure*}[t]
\includegraphics[width=0.97\textwidth]{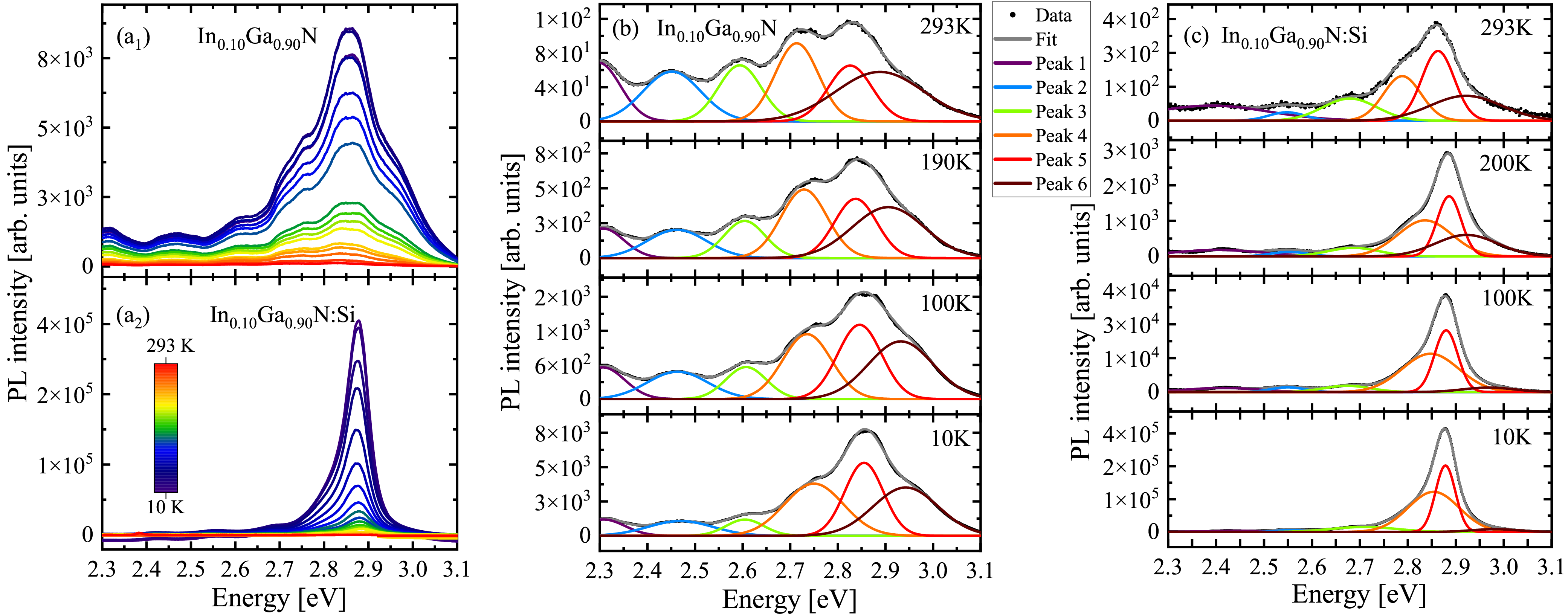} 
\caption{\label{fig:tempseries}The PL spectra showing in (a$_1$) and (a$_2$) the full temperature range for In$_{0.10}$Ga$_{0.90}$N and In$_{0.10}$Ga$_{0.90}$N:Si, respectively. (b) and (c) illustrate the representative Gaussian fitted spectra at selected temperatures for each sample, in that order.}
\end{figure*}

Silicon is a donor in ternary nitrides, hence it increases the electron concentration\cite{li2006influence,li2006enhanced,lu2013influence}. A Si doping of 0.5 at.\%, corresponding to carrier concentration of $\approx2.2\times10^{20}$ cm$^{-3}$, was used to achieve In$_{0.10}$Ga$_{0.90}$N:Si. The concentration exceeds the critical concentration, $n_c \approx 8\times10^{17}$ cm$^{-3}$ in In$_{0.10}$Ga$_{0.90}$N set by the Mott criterion\cite{Mott1990} for doped semiconductors given by $n^{1/3}_c a^{*}_{B} \approx 0.25$, where $a^{*}_{B}$ is the effective Bohr radius of the impurity atom, Si. This confirms a degenerate (metallic) system in Si-doped In$_{0.10}$Ga$_{0.90}$N as observed in Al$_{x}$Ga$_{1-x}$N\cite{Bharadwaj2019,Nishikawa2023}. Moreover, the high electron density entails a greater number of electrons available, which reduces the electron-hole separation and thereby increases wavefunction overlap, enhancing radiative recombination. Furthermore, this results in band-filling effects, known as the Burstein-Moss shift\cite{Burstein1954,Moss1954}, pushing the quasi-Fermi level higher in energy, modifying the recombination dynamics. The blueshift in the peak energies in the Si-doped sample, which is about 20 meV at 10 K, confirms the increased electron density due to doping. This counteracts the bandgap lowering that occurs due to the introduction of indium\cite{yamamoto2013}. Thus, the Si-doped In$_{0.10}$Ga$_{0.90}$N is degenerately doped and also optically semiconducting based on its enhanced luminescence properties.

The indium content in In$_{0.10}$Ga$_{0.90}$N is expected to cause alloy disorder, resulting from statistical variation in the distribution of In atoms within the InGaN lattice. This creates localized states that have been extensively studied in various conventional semiconducting materials\cite{baranovski2006charge,Weisbuch2021,baranovskii2022energy,David-Weisbusch2022} and, more recently, in low-dimensional systems\cite{xie2017inhomogeneous,Masenda2021}. Meanwhile, Si doping in In$_{0.10}$Ga$_{0.90}$N alloys is likely to introduce impurity disorder alongside In compositional fluctuation. Structurally, incorporating an impurity atom affects the lattice strain, as well as structural defects such as vacancies, interstitials, and dislocations, depending on the doping level. Structural analysis with XRD confirms lattice expansion due to alloying. This is evident from the increases in lattice parameters compared to pure GaN. Additionally, Si doping does not affect the lattice constants, indicating successful dopant incorporation, but it significantly increases edge dislocation densities, while screw dislocations remain unchanged. As a result, both alloying and doping are bound to influence the optical properties and resulting spectral features of In$_{0.10}$Ga$_{0.90}$N and In$_{0.10}$Ga$_{0.90}$N:Si. Over the years, several reports\cite{cho1998influence,cho1999influence,cheng2003quantum,wang2006band,lee2010} have shown that Si-doping in InGaN enhances crystal quality by reducing strain and suppressing vacancy-type defects. Thus, the In compositional variations (alloy disorder) in the In$_{0.10}$Ga$_{0.90}$N lattice lead to significant nonradiative recombination, which limits PL intensity and broadens spectral features. Yet, Si doping decreases defect density and helps passivate defects and alloy disorder, thus reducing nonradiative recombination channels. This favors radiative recombination, resulting in the observed enhanced PL intensity and narrower emission spectral linewidths. A deeper understanding of the nature and impact of induced localized states is explored by investigating the impact of temperature on luminescence properties, and this is discussed in the next section.

\subsection{Carrier localization and thermalization}

The nature of disorder induced by either alloying and/or doping can be explored from the temperature dependence of spectral features. The full temperature range PL spectra obtained for In$_{0.10}$Ga$_{0.90}$N and In$_{0.10}$Ga$_{0.90}$N:Si are shown in Fig. \ref{fig:tempseries}(a$_1$) and (a$_2$), respectively. Selected spectra at key equivalent temperatures in Fig. \ref{fig:tempseries}(b) and (c) illustrate how the peaks were deconvoluted using a Gaussian line shape to identify individual emission components. The most dominant emission line (peak 5) is attributed to near-band-edge emission, characteristic of a donor-bound exciton (D$^{0}$X), which forms when free excitons are trapped at a neutral donor site. Similarly, Nakamura et al.\cite{nakamura1992high} identified a comparable peak for In$_x$Ga$_{1-x}$N (with $x$ = 0.14 - 0.26) that also exhibited blue emission, with a linewidth of 70 - 110 meV, linked to near-band-edge transitions. The higher emission energy line, Peak 6, corresponds to free exciton A (FX$_{\rm {A}}$), typically seen in GaN\cite{xu2002direct,callsen2019probing}. The B excitons (FX$_{\rm {B}}$), usually at a much higher emission energy, are not expected as these are quenched for higher indium content such as  In$_{0.10}$Ga$_{0.90}$N\cite{callsen2019probing}. Additionally, the lower emission energy lines, Peaks 1 - 4, are characteristic of phonon-assisted transitions due to their broader linewidths compared to Peak 5. This wider linewidth signifies strong electron-phonon coupling, which is typically observed with longitudinal optical (LO) phonon replicas\cite{reshchikov2005luminescence}. Furthermore, the energy separation between the secondary peaks and the main one ranges from 88 to 137 meV. The optical phonon energy for InN and GaN is reported to be between 86 and 91 meV\cite{cao2019carrier}, while in InGaN/GaN quantum wells, the values can reach up to 105 meV\cite{feng2002phonon}. Moon et al.\cite{moon1999optical} observed spectral changes, including shifts in energy and the emergence of additional peaks, when PL was measured as a function of thickness. This indicates that the quantum well value does not completely explain the spectral features associated with thickness in thin films. Similar emission lines are observed for the Si-doped material, showing a blueshift when compared to the pristine In$_{0.10}$Ga$_{0.90}$N. As a result, the same assignments are adopted. As depicted in Fig. \ref{fig:InGaN}, the D$^{0}$X line is the most intense with a narrow linewidth, while the phonon replicas are significantly quenched. The temperature-dependent evolution of spectral features provides insight into the nature of disorder caused by alloying and/or doping. To highlight the thermal evolution of spectral features, the selected fitted spectra at equivalent temperatures are illustrated in Fig. \ref{fig:tempseries}(b) and (c). The D$^{0}$X (peak 5) in In$_{0.10}$Ga$_{0.90}$N undergoes a strong quenching at high temperatures, while the phonon replica components and FX$_{\rm {A}}$ are more pronounced. On the other hand, in In$_{0.10}$Ga$_{0.90}$N:Si, the doping promotes channels for recombination, hence the stability of the intensity of D$^{0}$X while suppressing that of the side features, including the free excitons (FX$_{\rm {A}}$). Thus confirming efficient radiative recombination from the localized states.

\begin{figure}[h]
\includegraphics[width=0.47\textwidth]{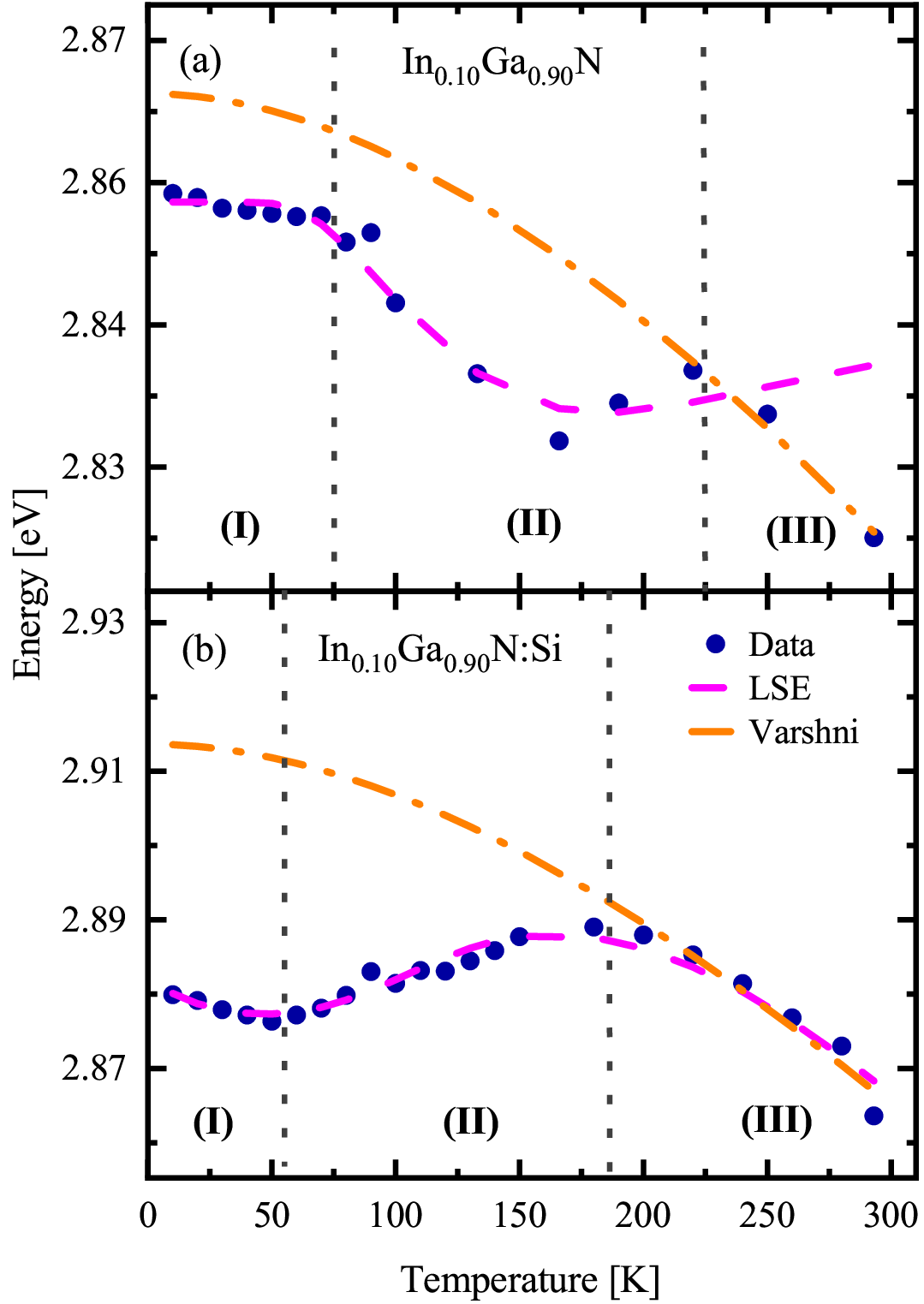}
\caption{\label{fig:energyfeat} Extracted peak energy positions against temperature fitted with the LSE model and Varshni's equation for (a) In$_{0.10}$Ga$_{0.90}$N and (b) In$_{0.10}$Ga$_{0.90}$N:Si}
\end{figure}

The peak energies for the D$^{0}$X extracted from Gaussian fits (10 - 293 K) are shown in Fig. \ref{fig:energyfeat}. For In$_{0.10}$Ga$_{0.90}$N (Fig. \ref{fig:energyfeat}(a)), the temperature dependence can be broken down into three regions: (I) 10 - 70 K, (II) 80 - 220 K, and (III) from 220 K to RT. In region (I), the energy remains nearly constant. In region (II), a rapid redshift occurs up to 165 K, followed by a blueshift. This redshift-blueshift behavior results from thermal escape and thermalization of localized carriers, a phenomenon common in semiconducting alloys such as InGaN\cite{wang2022analytic}. In the third region (III), the peak positions follow the usual band-gap shrinkage caused by lattice expansion at higher temperatures. 

As shown in Fig. \ref{fig:energyfeat}(b), the temperature-dependent PL peak position of the doped sample exhibits an $S$-shaped behavior, characteristic of carrier localization - a signature of potential fluctuations most likely stemming from impurity-induced disorder. Similarly, this plot can also be split into three regions: (I) 10 - 50 K, (II) 60 - 180 K, and (III) 200 K to RT. Region (I) shows a redshift from carrier activation within the disorder potential. A blueshift follows this in (II), which arises due to the carriers occupying higher energy states near the band-edge. Lastly, the third section again reflects a typical band gap shrinkage as observed in In$_{0.10}$Ga$_{0.90}$N, but at a lower temperature. It is worth noting that thermalization occurs at different temperatures: $\sim$165 K for the undoped and $\sim$50 K for the doped sample. This suggests different disorder origins. The former is due to alloy-induced indium fluctuations, resulting in inhomogeneity. The latter can be attributed to Si doping, which suppresses the fluctuation effect but introduces impurities that form deeper trap states. Further, over the measured temperature range, the blueshift in peak energy values of D$^{0}$X ranges from 20 - 50 meV, confirming the blueshift found in literature\cite{ryu2000silicon,cheng2003mechanisms,yamamoto2013}.

The alloy disorder in In$_{0.10}$Ga$_{0.90}$N generates a range of shallow potential minima (``localized states") that capture electrons and holes. These localized states spatially confine carriers, reducing their mobility and preventing them from recombining. In addition, the disorder is extensive, causing significant inhomogeneous broadening resulting in broad PL linewidths. On the other hand, Si doping introduces impurity-induced disorder, which leads to deeper and more well-defined localized states. This stronger localization further restricts carriers and reduces their overlap with nonradiative centers. Thus, the dopant-induced states are more discrete in energy and spatially confined, yielding sharper PL emission lines and enhanced radiative recombination. The findings from the XRD analysis of In$_{0.10}$Ga$_{0.90}$N samples provide valuable insights into how silicon (Si) doping affects dislocation densities. Specifically, the screw dislocation density remains consistent, while edge dislocation density increases by about 24\% with Si doping. This observation is especially relevant when considered alongside established theoretical frameworks\cite{Ourmazd1984,You2009} about how dislocations influence semiconductor properties. Screw dislocations primarily induce localized states through their associated strain fields, as discussed in the literature \cite{Albrecht2014,You2009}. Importantly, these screw dislocations can introduce both deep and shallow states, and their specific energy levels depend on the core structure and size in the particular semiconducting material. Conversely, the increase in edge dislocation density due to Si doping indicates that edge dislocations create deep localized states through their dangling bonds, effectively acting as traps for charge carriers\cite{You2009,Mishra2011}. The relationship between edge dislocation density and trap formation is important, as the XRD results show the impact of Si doping on defect structures in InGaN.

Thermal effects on the peak energies for the D$^{0}$X in undoped and doped In$_{0.10}$Ga$_{0.90}$N films were analyzed using Varshni's equation\cite{Varshni1967}. An empirical expression that accounts for band gap shrinkage with increasing temperature, which is given by,
\begin{equation}
    E(T) = E_0' - \frac{\alpha T^2}{\beta + T},
\end{equation}
\noindent where $E(T)$ is the energy at temperature $T$, $E_0'$ is the bandgap energy at 0 K, $\alpha$ and $\beta$ are Varshni's fitting parameters. Usually for InGaN, $\beta$ is the Debye temperature, which was calculated with Vegard's law for 10\% indium content\cite{gedam2017built}. The values obtained for the different variables are outlined in Table \ref{tab:Varshni}.

\begin{table}[h]
\caption{Parameters of Varshni's fit for the two samples}
    \centering
    \begin{tabular}{p{3cm} p{1.5cm} p{2cm} p{1.5cm}}
    \hline
         & $E'_0$ (eV) & $\alpha$ (meV/K) & $\beta$ (K) \\
    \hline
       In$_{0.10}$Ga$_{0.90}$N & 2.864 & 0.406 & 648.3 \\
       \hline
       In$_{0.10}$Ga$_{0.90}$N:Si & 2.910 & 0.486 & 648.3 \\
       \hline
    \end{tabular}
    \label{tab:Varshni}
\end{table}

As reflected in Fig. \ref{fig:energyfeat}, in both the alloy and after Si doping, Varshni's law is not sufficient to describe the low temperature dependence of the energy peak positions. Thus, the principle was applied at high temperatures. Consequently, to understand the nature of disorder in detail, the localized state ensemble (LSE) model by Li and Xu et al.\cite{li2001thermal,li2005model,li2005origin} initially employed for quantum wells was used. According to the model, the unusual temperature dependence of PL peak positions arises from the thermal redistribution of carriers within these localized states. The LSE model assumes a Gaussian-type density of localized states:

\begin{equation}\label{peak}
    E(T) = E_0 - \frac{\alpha T^2}{\beta + T} - x(T)k_B T, 
\end{equation}

\noindent where the temperature-dependent term $x(T)$ is defined by:

\begin{equation}\label{eqn:carrier}
     xe^x = [(\frac{\sigma}{k_B T})^2 -x](\frac{\tau_r}{\tau_{tr}})e^{(E_0-E_a)/k_B T},
\end{equation}

\noindent with $\sigma$ and $E_0$ representing the linewidth and center of the Gaussian distribution function for the localized states ($\rho = \rho_0 \text{exp}[-(E-E_0)/(2\sigma^2)]$, respectively. $E_a$ is the thermal escape energy, $\tau_r$ and $\tau_{tr}$ are the radiative recombination and transfer lifetimes, in that order. The extracted parameters from numerically fitting Equations \ref{peak} and \ref{eqn:carrier} are summarized in Table \ref{tab:energypara}.

\begin{table}[h]
\caption{Parameters of the LSE model for the two samples}
    \centering
    \begin{tabular}{p{2.5cm} p{1.3cm} p{1.5cm} p{1.4cm} p{1.1cm}}
    \hline
         & $E_0$ (eV) & $E_a$ (eV) &  $\sigma$ (meV) & $\tau_r/ \tau_{tr}$  \\
    \hline
       In$_{0.10}$Ga$_{0.90}$N & 2.854 & 2.908 &  18.96 & 326.9 \\
       \hline
       In$_{0.10}$Ga$_{0.90}$N:Si & 2.947 & 2.881  & 24.86 & 1.834  \\
       \hline
    \end{tabular}
    \label{tab:energypara}
\end{table}

The LSE model highlights the importance of $\Delta E = E_0 - E_a$ in determining the temperature dependence of the luminescence peak, particularly at low temperatures. Different notations are used in literature to define $\Delta E$, and that affects the sign. In this report, the energy difference is maintained as in Equation \ref{eqn:carrier}. Mohmad et al.\cite{mohmad2014localization} demonstrated how variations in this energy difference influence the peak positions. Here, $E_0$ is the center of the Gaussian distribution function, while $E_a$ depends on carrier concentrations and the material's internal electric field\cite{li2005origin}. Thus, $\Delta E$ provides the extensive energy related to the activation potential barrier and spatial distribution related to localized states or traps. As highlighted, $\Delta E$ may be positive or negative; each case has a physical significance. In disordered semiconductor systems, whether alloyed or doped, the energy levels for localized states are found within a complex potential landscape. The disorder related to alloying or impurities creates spatially varying potential wells and barriers instead of uniform conduction and valence bands, as found in a crystalline lattice. Electrons and holes are localized in these wells at specific energies $E_0$, which correspond to minima in the potential landscape, with an activation energy $E_a$ representing the energy the carrier must acquire to escape from that local potential trap into an extended or less localized state. The absolute values of $E_0$ and $E_a$ are measured relative to different local references within the traps, rather than a fixed global band edge, due to the spatial variation of the potential wells and barriers caused by the nature of the induced disorder. Consequently, this variation allows for situations where $E_0$ can be higher than $E_a$ when considering local potential energy, even though $E_a$ represents an energy barrier that the carrier must overcome to occupy higher energy levels. In$_{0.10}$Ga$_{0.90}$N:Si has $\Delta E$ = +66.11 meV, ($E_0$ $>$ $E_a$). Thus, carriers in impurity-induced localized states are deeply trapped in potential wells and must overcome an energy barrier to escape or recombine. As a result, the thermal activation barrier keeps the carriers trapped for longer times, resulting in a more pronounced carrier localization as reflected in the larger disorder parameter ($\sigma$). In contrast, for In$_{0.10}$Ga$_{0.90}$N, $\Delta E$ = -53.85 meV indicating $E_0$ $<$ $E_a$. In this case, the localized state lies below the effective barrier in the energy landscape. As a result, the carriers encounter relatively smaller or no energy barriers to escape, suggesting shallow trapping. Thus, the transition to higher levels is thermally favorable. For In$_{0.10}$Ga$_{0.90}$N:Si, as temperature increases, carriers gradually localize into deeper energy states, favoring transitions to lower-energy sites until reaching a minimum around 50 K. Beyond this point, thermal energy enables carriers to redistribute and overcome the $E_a$ barrier and occupy higher states, peaking around 180 K. Further temperature increase leads to lattice expansion and consequently the band-gap shrinkage. Conversely, for undoped In$_{0.10}$Ga$_{0.90}$N, carriers are initially in shallow and spatially distributed localized states. This results in a nearly constant emission energy below 70 K. From 80 K to 165 K, the localized states induced by the alloy disorder are energetically favorable, resulting in carriers accessing deeper traps, reaching a minimum at 165 K. Beyond this, they begin occupying higher-energy states until about 225 K, after which the band gap narrows due to lattice expansion. The $S$-shaped behavior is dominated by thermal redistribution within a broad disordered landscape rather than only overcoming the activation barrier. 

Therefore, the sign and magnitude of $\Delta E$ ($E_0 - E_a$) indicate how strongly carriers are localized. These also relate to the difficulty of thermal escape in a disorder-induced potential, either by alloying or incorporating foreign atoms. Thus, Si doping increases the degree of localization as reflected in the increase in $\sigma$ from 18.96 meV to 24.87 meV, indicating deeper trap states. The relatively larger disorder parameter coupled with a positive energy difference $\Delta E$ confirms that Si introduces impurity states and modifies the local atomic arrangements due to alloy disorder in the In$_{0.10}$Ga$_{0.90}$N. Moreover, the Stokes shift, the energy difference between the band gap according to Varshni's law and the PL peak, is much larger in In$_{0.10}$Ga$_{0.90}$N:Si when compared to In$_{0.10}$Ga$_{0.90}$N  at $T < 140$ K. In the range, $150 < T < 200$ K, the Stokes shift for the alloy is slightly higher, which coincides with the region when the alloy disorder is more prominent.  Lastly, at higher temperatures, the peak energies in both materials follow the shrinkage of the optical bandgap with increasing temperature.

As highlighted in the preceding paragraphs, the observed differences in luminescence spectral features between pristine and Si-doped In$_{0.10}$Ga$_{0.90}$N alloys result from the intricate interplay of two main competing effects: recombination processes (radiative/nonradiative) and carrier localization. Furthermore, another key competing process between the pristine alloy and the Si-doped is the thermalization of carriers, which is at play. This is evident in different $S$-shape behaviors. As observed in Fig. \ref{fig:energyfeat}(a) and (b), the redshift-to-blueshift transition temperature changes from around 165 K in In$_{0.10}$Ga$_{0.90}$N films down to about 50 K after Si doping. In In$_{0.10}$Ga$_{0.90}$N, with increasing temperature, carriers trapped in shallow alloy fluctuation states start populating deeper energetically favorable sites around 80 K.  After which, the carriers hop or redistribute among localized states, causing an initial redshift in PL peak energy reaching transition at $\sim$ 165 K which is then followed by a blueshift (thermal escape to higher energy sites). Surprisingly, the deeper localized states created by the incorporation of Si require a lower thermal activation energy for redistribution, such that carrier thermalization and hopping occur at a lower temperature ($\sim$50 K), which seems contradictory. However, although the states are deeper, they are energetically more discrete. This allows carrier rearrangement or hopping to even deeper localized states at lower temperatures, which reduces the $S$-shape transition after Si doping. Despite being deeper, impurity-induced localized states in Si-doped samples are energetically well-separated and more spatially confined, which enhances the coupling between localized states. This configuration leads to more efficient thermalization and carrier redistribution at lower temperatures; carriers trapped in these deep states require less thermal energy to hop or tunnel between energy levels. In a way, Si doping creates deeper energy states, resulting in distinct localized traps with sharper potential wells, which can easily saturate. As the temperature increases, the saturation prevents continuous carrier relaxation to even deeper states, effectively forcing carriers to occupy higher energy sites at lower temperatures than in In$_{0.10}$Ga$_{0.90}$N. This facilitates the thermal activation of carrier transfer and redistribution at lower temperatures. On the other hand, the alloy fluctuations introduced by 10\% In concentration yield a wider spread or more spatially isolated shallower states that reduce the coupling and hopping of carriers between localized states, requiring higher temperatures for significant carrier thermalization and redistribution to occur. 

\subsection{Carrier dynamics}
\begin{figure*}[t]
\includegraphics[width=0.99\textwidth]{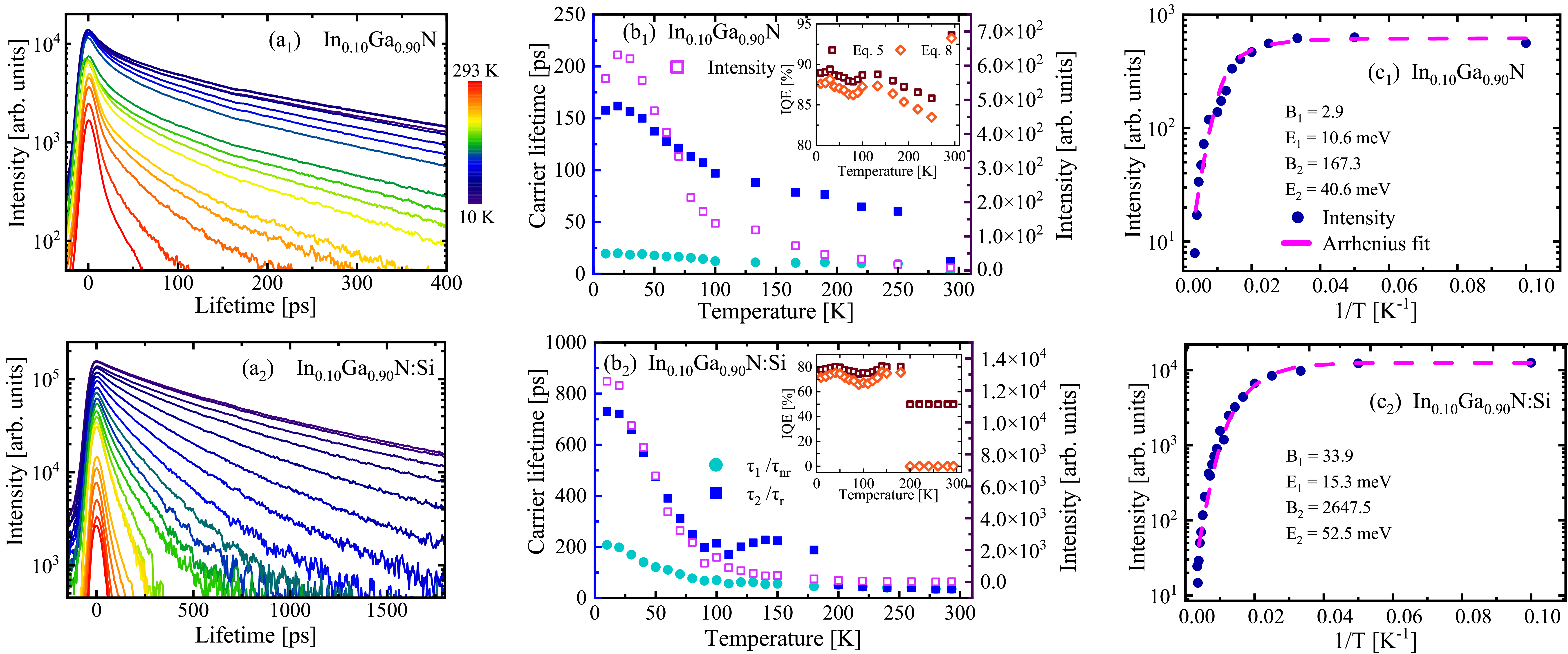}
\caption{\label{fig:time}The temporal evolution of the transient spectra and intensity of In$_{0.10}$Ga$_{0.90}$N (top) and {In$_{0.10}$Ga$_{0.90}$N:Si} (bottom). Shown in (a$_1$) and (a$_2$) are the spectra obtained from TD-TRPL measurements, (b$_1$) and (b$_2$) depict the extracted $\tau_1$ and $\tau_2$ parameters for the samples, including their respective intensity curves. Lastly, the fitted thermal quenching of the PL intensity in (c$_1$) and (c$_2$).}
\end{figure*} 

As clearly observed in Fig. \ref{fig:energyfeat}(a), the LSE model is sufficient to describe the temperature dependence of the peak energy position in In$_{0.10}$Ga$_{0.90}$N at $T < 200$ K but fails thereafter. This is also reflected in the large $\tau_r/\tau_{tr}$ ratio. To further examine carrier recombination dynamics, the temporal components of the time-resolved PL data obtained between 10 K and 293 K were analyzed. The transient spectra for In$_{0.10}$Ga$_{0.90}$N and In$_{0.10}$Ga$_{0.90}$N:Si across the entire temperature range are shown in Fig. \ref{fig:time}(a$_1$) and (a$_2$), respectively. The spectra were fitted with a bi-exponential function given by, 
\begin{align}\label{eqn:time}
    I(t) &= I_0 + A_1e^{-\frac{t}{\tau_1}} + A_2e^{-\frac{t}{\tau_2}},
\end{align}
where $I_0$ is the initial PL intensity, $t$ is the time, $\tau_1$ and $\tau_2$ are the fast and slow carrier lifetimes, in that order. The fast decay represents the nonradiative recombination lifetime ($\tau_{nr}$), while the slow decay presents the radiative recombination lifetime ($\tau_{r}$)\cite{thaalbi2025streamlined}. The extracted parameters are presented in Fig. \ref{fig:time}(b$_1$) and (b$_2$), respectively. Evidently, both the radiative and nonradiative recombination lifetimes for In$_{0.10}$Ga$_{0.90}$N:Si are longer. This observation confirms that Si doping introduces deeper localized states. These states make recombination slower compared to the shallower energy levels in In$_{0.10}$Ga$_{0.90}$N. The radiative recombination trend for In$_{0.10}$Ga$_{0.90}$N (Fig. \ref{fig:time}(b$_1$)) shows a pronounced decline at temperatures up to 100 K, followed by a much slower decrease in lifetime up to 250 K. Conversely, the nonradiative recombination for this sample displays an almost steady trend throughout the whole temperature range. In the In$_{0.10}$Ga$_{0.90}$N:Si plot (Fig. \ref{fig:time}(b$_2$)), the curve for radiative recombination exhibits a significantly steeper decline. There is a rapid decrease from 710 ps at 10 K to 200 ps at 90 K, followed by a further decline above 150 K. Furthermore, at temperatures $T > 200$ K, $\tau_1 \approx \tau_2$, indicating a transition from a bi-exponential function to mono-exponential dependence. This suggests the merging of different recombination channels. Additionally, the coincidence with the band-gap shrinkage (redshift) points to a strong temperature coupling between carrier dynamics and band structure fluctuations. The limitation with a mono-exponential decay is the inability to calculate the internal quantum efficiency (IQE) accurately. The conventional approach for determining the IQE relies on the ABC model, where A, B, and C represent the coefficients for radiative, non-radiative, and Auger recombination, respectively. Auger recombination occurs when an electron and a hole recombine, instead of releasing a photon, a carrier is excited to a higher energy state\cite{delaney2009auger}. In systems with higher carrier densities, Auger recombination is expected to increase. However, the Auger coefficient is generally quite small in InGaN, even at higher carrier concentrations\cite{shim2013active,yu2022invalidation}. Therefore, its impact is often considered negligible due to its relatively minor contribution. Consequently, IQE is generally calculated using the AB model, which considers the effective PL lifetime ($\tau_{\text{PL}}$) as a sum of the radiative and non-radiative lifetimes. In this case, the $\tau_{\text{PL}}$ contributors are derived from a bi-exponential fit. Thus, the ratio used to establish IQE is,
\begin{align}
    \text{IQE} = \frac{\tau_{r}}{\tau_{nr}+\tau_{r}}.
    \label{eq:ratio}
\end{align}
The inset in Fig. \ref{fig:time}(b$_1$) has the IQE for In$_{0.10}$Ga$_{0.90}$N. The steady decrease in radiative recombination, while the nonradiative transition remains nearly constant for this sample, resulted in an IQE that averaged out to 89\% within the temperature range of 10 - 130 K. Beyond this point, the IQE slightly declines to 86\%. The inset in Fig. \ref{fig:time}(b$_2$) shows the IQE plot for In$_{0.10}$Ga$_{0.90}$N:Si. In comparison, this sample's IQE had an average of 78\% in the temperature range with a bi-exponential decay and a drop to 50\% in the $T \geq 200$ K.

Alternatively, another approach by Kim et al.\cite{kim2010analysis, kim2012estimate}, which is also based on the AB-model, albeit used primarily to estimate IQE at RT, was employed. For this method, the $\tau_{nr}$ and $\tau_{r}$ as well as the IQE can be approximated by:

\begin{align}
    \tau_{nr} &\simeq 2\tau_{slow}\\
    \tau_{r} &= \frac{2\tau_{slow}\tau_{fast}}{\tau_{slow}-\tau_{fast}} \\
    \text{IQE} &= \frac{\tau_{slow}-\tau_{fast}}{\tau_{slow}}\label{eqn:recom} 
\end{align}

\noindent Here, the individual transient spectra are approximated as single exponential functions for the fast (or initial) and slow (or final) decay stages for the different temperatures measured. The temperature dependence of the IQE is depicted in the insets in Fig. \ref{fig:time}(b$_1$) and (b$_2$). The IQE for In$_{0.10}$Ga$_{0.90}$N shows a fluctuating trend from 10 to 130 K, with an average of at least 87\%, followed by a decrease to 83\% at higher temperatures. This trend corresponds to the consistently short radiative recombination lifetimes within the same temperature range. Similarly, In$_{0.10}$Ga$_{0.90}$N:Si also has a fluctuating IQE trend, driven by the pronounced changes in radiative recombination lifetimes. The longer radiative times contribute to the slightly lower IQE observed in the Si-doped samples. PL intensity curves are included in Fig. \ref{fig:time}(b$_1$) and (b$_2$) to demonstrate the expected trend of the IQE with increasing temperature. This analysis reveals that both equations produced efficiencies that deviate from this. For the samples investigated, the equations displayed a similar trend, although Kim's model yielded a lower IQE. Further, as expected, Equation \ref{eqn:recom} proved insufficient to accurately describe the efficiency when the bi-exponential decay in In$_{0.10}$Ga$_{0.90}$N:Si transitioned into a mono-exponential form. Throughout that temperature range, the efficiency calculated dropped to zero, while the other equation showed a decrease to 50\%. It is also worth noting that Equation \ref{eqn:recom} is typically used for time scales longer than those in Fig. \ref{fig:time}, which may have further contributed to its limitations. 

The thermal quenching of the PL intensity is shown in Fig. \ref{fig:time}(c$_1$) and (c$_2$). The Arrhenius equation was used to fit the data,
\begin{align}
    I(T) = \frac{I_0}{1+B_1 e^{\frac{-E_1}{k_BT}}+B_2 e^{\frac{-E_2}{k_BT}}},
\end{align}
where $I_0$ represents maximum intensity, while $E_1$ and $E_2$ are thermal activation energies for nonradiative recombination centers. $B_1$ and $B_2$ are coefficients relating to the density of the nonradiative centers. The fitting for In$_{0.10}$Ga$_{0.90}$N yielded $E_1$ = 10.6 meV and $E_2$ = 40.6 meV, while In$_{0.10}$Ga$_{0.90}$N:Si resulted in $E_1$ = 15.3 meV and $E_2$ = 52.5 meV. $E_1$ corresponds to the thermal escape energy for weakly localized states, and $E_2$ for strongly localized states. The $E_2$ value for In$_{0.10}$Ga$_{0.90}$N:Si supports the LSE model findings, suggesting the presence of deeper trap states as a result of doping, which requires carriers to possess greater energy for thermal escape. Additionally, the significantly larger $B_2$ coefficient for In$_{0.10}$Ga$_{0.90}$N:Si ($B_2$ = 2647.5) compared to the undoped sample ($B_2$ = 167.3) further highlights the longer carrier lifetimes observed in Fig. \ref{fig:time}(b$_2$) in contrast to Fig. \ref{fig:time}(b$_1$).

Overall, the rocking curve and RSM analyses indicated that these epitaxially grown samples were high-quality crystals. The results also showed that the introduction of silicon had a negligible effect on the in-plane and out-of-plane lattice parameters. Dislocation analysis revealed that doping primarily influenced the edge dislocation. Nevertheless, the dislocation density remained within the expected range (mid-$10^9$ cm$^{-2}$) for good quality crystals suitable for optoelectronic applications. In addition, the PL measurements demonstrated that silicon alters disorder potential landscapes in an alloyed material system by passivating defects. It changed the nature of induced disorder, thereby introducing different recombination channels. These appeared as observable changes in time-resolved photoluminescence studies. The presence of the $S$-shape in both pristine and doped In$_{0.10}$Ga$_{0.90}$N films confirmed the existence of induced disorder potentials. These potentials have different origins, based on the observed luminescence properties and extracted parameters, along with their temperature dependence. Varshni's law adequately described the high-temperature variation of peak positions, allowing extrapolation of their behavior at cryogenic temperatures. Applying the LSE model enabled quantification of the impact of impurity and alloy disorder in the ternary nitride systems. This showed that the observed optical properties, carrier dynamics, and internal quantum efficiency were a result of a complex interplay between radiative and nonradiative recombination, carrier localization caused by either alloy or impurity-induced disorder, and their thermalization processes. Additionally, the doping level surpasses the estimated Mott critical concentration ($n_c\approx 10^{18}$). This suggests that the Si-doped InGaN is a degenerate (metallic) system characterized by a significantly higher density of free carriers and partial delocalization. Consequently, Burstein–Moss band filling effects are anticipated. Furthermore, the PL spectra reveal enhanced, blue-shifted donor-bound exciton emission, evidenced by narrower linewidths and reduced side peaks. The more pronounced $S$-shaped temperature dependence and the increase in disorder parameters following Si doping indicate a stronger localization within a disordered potential landscape. These findings imply a coexistence of an induced metallic (degenerate) transport properties along with preserved semiconducting and localization-driven optical behaviors in the Si-doped InGaN layer. This presents promising potential for applications in cladding layers for InGaN-based laser diodes, transparent electrodes, UV/blue photonic structures, waveguides, and photonic crystals.

\section{Conclusion}

Temperature-dependent TRPL studies on In$_{0.10}$Ga$_{0.90}$N revealed the presence of alloy disorder, while Si doping altered the optical properties and carrier dynamics in In$_{0.10}$Ga$_{0.90}$N. At 10 K, the doped sample showed enhanced blue-shifted emission lines with narrower linewidths and suppressed side-peaks; a trend that persisted up to 293 K. Temperature-dependent peak energy positions for the most dominant donor-bound exciton displayed non-monotonous dependence ($S$-shape) behavior in both samples. In In$_{0.10}$Ga$_{0.90}$N, this is due to alloy disorder, whereas in In$_{0.10}$Ga$_{0.90}$N:Si, it arises from impurity-induced disorder. Broad alloy fluctuations lead to shallow and spatially extended localized states. These states promote carrier redistribution at elevated temperatures and nonradiative processes, which diminish PL intensity. In contrast, Si doping passivated alloy disorder and introduced deeper, more discrete localized states. This enhances carrier confinement, leading to the early saturation of the localized states and shifting the onset of thermalization to approximately 50 K from 165 K for the undoped sample. The pronounced localization in In$_{0.10}$Ga$_{0.90}$N:Si effectively promotes radiative recombination, resulting in higher PL intensity and narrower emission linewidths. Even though XRD confirmed good crystal quality, alloy and impurity-induced disorder still prevailed. Alloying tunes the structural and optical properties, which is evident with the expansion of the lattice due to 10\% indium and lowering of the bandgap. On the other hand, Si doping does not affect lattice parameters, proving successful doping via MOVPE growth. Moreover, the introduction of Si only increases the edge dislocation density and the bandgap; it counteracts the redshift caused by alloying. This study offers valuable insights into the complexities of introducing Si into alloyed InGaN. It highlights the influences of various sources of disorder and concurrently induces a degenerate semiconductor, which is crucial for charge transport in optoelectronic applications.

\begin{acknowledgements}

O. Mpatani and H. Masenda acknowledge support from the South African National Research Foundation and the Department of Science and Innovation within the SA-CERN programme.  O. Mpatani is grateful for support from the PhD SA-CERN Excellence Bursary and the Wits Postgraduate Merit Award. D. Muth, A. Krüger, and M. Gerhard acknowledge support from the German Research Foundation DFG via the Collaborative Research Centre No. 223848855-SFB 1083, sub-projects B10, B12, and B15. H. Masenda acknowledges support from the Alexander von Humboldt (AvH) Foundation within the Georg Forster Research Fellowship program. R. Adhikari and A. Bonanni acknowledge funding from the Austrian Science Fund (FWF) through Projects No. P26830 and No. P31423. R. Adhikari also acknowledges the funding from the Austrian Science Fund (FWF) through Project No. TAI-817 and from the JKU LIT Seed funding through Project No. LIT-2022-11-SEE-119. 
\end{acknowledgements}

\onecolumngrid
\section*{Data Availability Statement}
\noindent The data that support the findings of this study are available from the corresponding authors upon reasonable request.

\section*{Conflict of interest}
\begin{center}
    The authors declare no conflict.
\end{center}
\onecolumngrid
\twocolumngrid
\onecolumngrid
\vspace{-2mm}
\section*{References}
\vspace{-5.5mm}
%\twocolumngrid
\vspace{-5.5mm}
\bibliography{ref}

@article{Anderson1958,
  title = {Absence of Diffusion in Certain Random Lattices},
  author = {Anderson, P. W.},
  journal = {Physical Review},
  volume = {109},
  issue = {5},
  pages = {1492--1505},
  numpages = {0},
  year = {1958},
  month = {Mar},
  publisher = {American Physical Society},
  doi = {https://doi.org/10.1103/PhysRev.109.1492}
}

@article{baranovskii2022energy,
  title={Energy scales of compositional disorder in alloy semiconductors},
  author={Baranovskii, Sergei D and Nenashev, Alexey V and Hertel, Dirk and Gebhard, Florian and Meerholz, Klaus},
  journal={ACS omega},
  volume={7},
  number={50},
  pages={45741--45751},
  year={2022},
  publisher={ACS Publications},
  doi = {https://doi.org/10.1021/acsomega.2c05426}
}

@article{nakamura1993si,
  title={{Si-doped InGaN films grown on GaN films}},
  author={Nakamura, Shuji and Mukai, Takashi Mukai Takashi and Senoh, Masayuki Senoh Masayuki},
  journal={Japanese Journal of Applied Physics},
  volume={32},
  number={1A},
  pages={L16},
  year={1993},
  publisher={IOP Publishing},
  doi = {https://doi.org/10.1143/JJAP.32.L16}
}

@article{nakamura1992high,
  title={{High-quality InGaN films grown on GaN films}},
  author={Nakamura, Shuji and Mukai, Takashi},
  journal={Japanese Journal of Applied Physics},
  volume={31},
  number={10B},
  pages={L1457},
  year={1992},
  publisher={IOP Publishing},
  doi = {https://doi.org/10.1143/JJAP.31.L1457}
}

@article{pampili2017doping,
  title={{Doping of III-nitride materials}},
  author={Pampili, Pietro and Parbrook, Peter J},
  journal={Materials Science in Semiconductor Processing},
  volume={62},
  pages={180--191},
  year={2017},
  publisher={Elsevier},
  doi = {https://doi.org/10.1016/j.mssp.2016.11.006}
}

@book{Adachi2009,
   author = {Sadao Adachi},
   city = {Chichester, UK},
   month = {3},
   publisher = {John Wiley \& Sons, Ltd},
   title = {{Properties of Semiconductor Alloys: Group-IV, III-V and II-VI Semiconductors}},
   year = {2009},
   doi = {https://doi.org/10.1002/9780470744383}
}

@book{Nakamura2000,
   author = {Shuji Nakamura and Shigefusa F. Chichibu},
   doi = {10.1201/9781482268065},
   journal = {Introduction to Nitride Semiconductor Blue Lasers and Light Emitting Diodes},
   month = {3},
   publisher = {CRC Press},
   title = {Introduction to Nitride Semiconductor Blue Lasers and Light Emitting Diodes},
   url = {https://www.taylorfrancis.com/books/9780429179556},
   year = {2000},
}

@book{Morkoc2009,
   author = {Hadis Morkoç},
   doi = {https://doi.org/10.1002/9783527628438},
   journal = {{Handbook of Nitride Semiconductors and Devices}},
   month = {3},
   publisher = {Wiley},
   title = {{Handbook of Nitride Semiconductors and Devices}},
   Volume = {1},
   url = {https://onlinelibrary.wiley.com/doi/book/10.1002/9783527628438},
   year = {2009},
}

@article{nakamura2015nobel,
  title={{Nobel Lecture: Background story of the invention of efficient blue InGaN light emitting diodes}},
  author={Nakamura, Shuji},
  journal={Reviews of Modern Physics},
  doi = {10.1103/RevModPhys.87.1139},
  volume={87},
  number={4},
  pages={1139},
  year={2015},
  publisher={APS}
}

@article{chen2020recent,
  title={{Recent progress in group III-nitride nanostructures: From materials to applications}},
  author={Chen, Fei and Ji, Xiaohong and Lau, Shu Ping},
  journal={Materials Science and Engineering: R: Reports},
  volume={142},
  pages={100578},
  year={2020},
  publisher={Elsevier},
  doi = {https://doi.org/10.1016/j.mser.2020.100578}
}

@article{moustakas2017optoelectronic,
  title={{Optoelectronic device physics and technology of nitride semiconductors from the UV to the terahertz}},
  author={Moustakas, T.D. and Paiella, Ro},
  journal={Reports on Progress in Physics},
  volume={80},
  doi = {10.1088/1361-6633/aa7bb2},   
  number={10},
  pages={106501},
  year={2017},
  publisher={IOP Publishing}
}

@article{reshchikov2005luminescence,
  title={{Luminescence properties of defects in GaN}},
  author={Reshchikov, Michael A and Morko{\c{c}}, Hadis},
  journal={Journal of Applied Physics},
  volume={97},
  number={6},
  year={2005},
  publisher={AIP Publishing},
  url = {https://doi.org/10.1063/1.1868059}
}

@article{li2006influence,
  title={{Influence of Si doping on the optical and structural properties of InGaN films}},
  author={Li, Da-Bing and Katsuno, Takuya and Nakao, Keisuke and Aoki, Masakazu and Miyake, Hideto and Hiramatsu, Kazumasa},
  journal={Journal of Crystal Growth},
  volume={290},
  number={2},
  pages={374--378},
  year={2006},
  publisher={Elsevier},
  doi = {https://doi.org/10.1016/j.jcrysgro.2006.01.058}
}

@article{gedam2017built,
  title={{Built-in-polarization and thermal conductivity of In$_x$Ga$_{1-x}$N/GaN heterostructures}},
  author={Gedam, V and Sahoo, BK},
  journal={Physica E: Low-dimensional Systems and Nanostructures},
  volume={93},
  pages={63--69},
  year={2017},
  publisher={Elsevier},
  doi = {https://doi.org/10.1016/j.physe.2017.04.024}
}

@article{lu2013influence,
  title={{Influence of Si doping on the structural and optical properties of InGaN epilayers}},
  author={Lu, Ping-Yuan and Ma, Zi-Guang and Su, Shi-Chen and Zhang, Li and Chen, Hong and Jia, Hai-Qiang and Jiang, Yang and Qian, Wei-Ning and Wang, Geng and Lu, Tai-Ping and others},
  journal={Chinese Physics B},
  volume={22},
  number={10},
  pages={106803},
  year={2013},
  publisher={IOP Publishing},
  doi = {10.1088/1674-1056/22/10/106803}
}

@article{li2005model,
  title={{A model for steady-state luminescence of localised-state ensemble}},
  author={Li, Q and Xu, SJ and Xie, MH and Tong, SY},
  journal={Europhysics Letters},
  volume={71},
  number={6},
  pages={994},
  year={2005},
  publisher={IOP Publishing},
  doi = {10.1209/epl/i2005-10170-7}
}

@article{li2005origin,
  title={{Origin of the ‘S-shaped’ temperature dependence of luminescent peaks from semiconductors}},
  author={Li, Q and Xu, SJ and Xie, MH and Tong, SY},
  journal={Journal of Physics: Condensed Matter},
  volume={17},
  number={30},
  pages={4853},
  year={2005},
  publisher={IOP Publishing},
  doi = {10.1088/0953-8984/17/30/011}
}

@article{li2001thermal,
  title={{Thermal redistribution of localised excitons and its effect on the luminescence band in InGaN ternary alloys}},
  author={Li, Q and Xu, SJ and Cheng, WC and Xie, MH and Tong, SY and Che, CM and Yang, H},
  journal={Applied Physics Letters},
  volume={79},
  number={12},
  pages={1810--1812},
  year={2001},
  publisher={American Institute of Physics},
  doi = {https://doi.org/10.1063/1.1403655}
}

@article{nakamura2013history,
  title={{History of gallium--nitride-based light-emitting diodes for illumination}},
  author={Nakamura, Shuji and Krames, Michael R},
  journal={Proceedings of the IEEE},
  volume={101},
  number={10},
  pages={2211--2220},
  year={2013},
  publisher={IEEE},
  doi = {https://doi.org/10.1109/JPROC.2013.2274929}
}

@article{hardy2011group,
  title={{Group III-nitride lasers: A materials perspective}},
  author={Hardy, Matthew T and Feezell, Daniel F and DenBaars, Steven P and Nakamura, Shuji},
  journal={Materials Today},
  volume={14},
  number={9},
  pages={408--415},
  year={2011},
  publisher={Elsevier},
  doi = {https://doi.org/10.1016/S1369-7021(11)70185-7}
}

@article{cai2023red,
  title={{Red emission from InGaN active layer grown on nanoscale InGaN pseudosubstrates}},
  author={Cai, Wentao and Wang, Jia and Park, Jeong-Hwan and Furusawa, Yuta and Cheong, Heajeong and Nitta, Shugo and Honda, Yoshio and Pristovsek, Markus and Amano, Hiroshi},
  journal={Japanese Journal of Applied Physics},
  volume={62},
  number={2},
  pages={020902},
  year={2023},
  publisher={IOP Publishing},
 doi = {10.35848/1347-4065/acb74c}
}

@article{funato2013remarkably,
  title={{Remarkably suppressed luminescence inhomogeneity in a (0001) InGaN green laser structure}},
  author={Funato, Mitsuru and Kim, Yoon Seok and Hira, Takayuki and Kaneta, Akio and Kawakami, Yoichi and Miyoshi, Takashi and Nagahama, Shin-Ichi},
  journal={Applied Physics Express},
  volume={6},
  number={11},
  pages={111002},
  year={2013},
  publisher={IOP Publishing},
  doi = {10.7567/APEX.6.111002}
}

@article{ryu2001effects,
  title={{Effects of Si-doping in the barriers on the recombination dynamics in In$_{0.15}$Ga$_{0.85}$N/In$_{0.015}$Ga$_{0.985}$N quantum wells}},
  author={Ryu, Mee-Yi and Yu, Phil Won and Shin, Eun-joo and Lee, Joo In and Yu, Sung Kyu and Oh, Eunsoon and Nam, Ok Hyun and Sone, Chul Soo and Park, Yong Jo},
  journal={Journal of Applied Physics},
  volume={89},
  number={1},
  pages={634--637},
  year={2001},
  publisher={American Institute of Physics},
  doi = {https://doi.org/10.1063/1.1331077}
}

@article{bose2018imaging,
  title={{Imaging localised energy states in silicon-doped InGaN nanowires using 4D electron microscopy}},
  author={Bose, Riya and Adhikari, Aniruddha and Burlakov, Victor M and Liu, Guangyu and Haque, Md Azimul and Priante, Davide and Hedhili, Mohamed N and Wehbe, Nimer and Zhao, Chao and Yang, Haoze and others},
  journal={ACS Energy Letters},
  volume={3},
  number={2},
  pages={476--481},
  year={2018},
  publisher={ACS Publications},
  doi = {https://doi.org/10.1021/acsenergylett.7b01330}
}

@article{cheng2003mechanisms,
  title={{Mechanisms for photon-emission enhancement with silicon doping in InGaN/GaN quantum-well structures}},
  author={Cheng, Yung-Chen and Tseng, Cheng-Hua and Hsu, Chen and Ma, Kung-Jen and Feng, Shih-Wei and Lin, En-Chiang and Yang, CC and Chyi, Jen-Inn},
  journal={Journal of Electronic Materials},
  volume={32},
  pages={375--381},
  year={2003},
  publisher={Springer},
  doi = {https://doi.org/10.1007/s11664-003-0161-8}
}

@article{wang2022analytic,
  title={{Analytic S-Shaped temperature dependence of peak positions of the localised-state ensemble luminescence and application in the analysis of luminescence in non-and semi-polar InGaN/GaN quantum-wells micro-array}},
  author={Wang, Xiaorui and Xu, Shijie},
  journal={Chinese Physics Letters},
  volume={39},
  number={10},
  pages={107801},
  year={2022},
  publisher={IOP Publishing},
  doi = {10.1088/0256-307X/39/10/107801}
}

@article{chichibu2006origin,
  title={{Origin of defect-insensitive emission probability in In-containing (Al,In,Ga)N alloy semiconductors}},
  author={Chichibu, Shigefusa F and Uedono, Akira and Onuma, Takeyoshi and Haskell, Benjamin A and Chakraborty, Arpan and Koyama, Takahiro and Fini, Paul T and Keller, Stacia and DenBaars, Steven P and Speck, James S and others},
  journal={Nature Materials},
  volume={5},
  number={10},
  pages={810--816},
  year={2006},
  publisher={Nature Publishing Group UK London},
  doi = {https://doi.org/10.1038/nmat1726}
}

@article{ponce2003microstructure,
  title={{Microstructure and electronic properties of InGaN alloys}},
  author={Ponce, FA and Srinivasan, S and Bell, A and Geng, L and Liu, R and Stevens, M and Cai, J and Omiya, H and Marui, H and Tanaka, S},
  journal={Physica Status Solidi (b)},
  volume={240},
  number={2},
  pages={273--284},
  year={2003},
  publisher={Wiley Online Library},
  doi = {https://doi.org/10.1002/pssb.200303527}
}

@article{lu2014temperature,
  title={{Temperature-dependent photoluminescence in light-emitting diodes}},
  author={Lu, Taiping and Ma, Ziguang and Du, Chunhua and Fang, Yutao and Wu, Haiyan and Jiang, Yang and Wang, Lu and Dai, Longgui and Jia, Haiqiang and Liu, Wuming and others},
  journal={Scientific Reports},
  volume={4},
  number={1},
  pages={6131},
  year={2014},
  publisher={Nature Publishing Group UK London},
  doi = {https://doi.org/10.1038/srep06131}
}

@article{li2016time,
  title={{Time-resolved photoluminescence studies of InGaN/GaN multi-quantum-wells blue and green light-emitting diodes at room temperature}},
  author={Li, Qiang and Wang, Shuai and Gong, Zhi-Na and Yun, Feng and Zhang, Ye and Ding, Wen},
  journal={Optik},
  volume={127},
  number={4},
  pages={1809--1813},
  year={2016},
  publisher={Elsevier},
  doi = {https://doi.org/10.1016/j.ijleo.2015.11.095}
}

@article{pecharroman2004investigation,
  title={{Investigation of the unusual temperature dependence of InGaN/GaN quantum well photoluminescence over a range of emission energies}},
  author={Pecharrom{\'a}n-Gallego, R and Martin, RW and Watson, IM},
  journal={Journal of Physics D: Applied Physics},
  volume={37},
  number={21},
  pages={2954},
  year={2004},
  publisher={IOP Publishing},
  doi = {10.1088/0022-3727/37/21/003}
}

@article{kim2012estimate,
  title={{Estimate of the nonradiative carrier lifetime in InGaN/GaN quantum well structures by using time-resolved photoluminescence}},
  author={Kim, Hyunsung and Han, Dong-Pyo and Oh, Ji-Yeon and Shim, Jong-In and Shin, Dong-Soo and Ryu, Han-Youl},
  journal={Journal of the Korean Physical Society},
  volume={60},
  number={11},
  pages={1934--1938},
  year={2012},
  publisher={Springer},
  doi = {https://doi.org/10.3938/jkps.60.1934}
}

@article{feng2002phonon,
  title={{Phonon--replica transitions in InGaN/GaN quantum well structures}},
  author={Feng, S-W and Tsai, C-Y and Cheng, Y-C and Liao, C-C and Yang, CC and Lin, Y-S and Ma, K-J and Chyi, J-I},
  journal={Optical and Quantum Electronics},
  volume={34},
  number={12},
  pages={1213--1219},
  year={2002},
  publisher={Springer},
  doi = {https://doi.org/10.1023/A:1021382721938}
}

@article{li2006enhanced,
  title={{Enhanced emission efficiency of InGaN films with Si doping}},
  author={Li, Da-Bing and Liu, Yu-Huai and Katsuno, Takuya and Nakao, Keisuke and Nakamura, Kazuya and Aoki, Masakazu and Miyake, Hideto and Hiramatsu, Kazumasa},
  journal={Physica Status Solidi C},
  volume={3},
  number={6},
  pages={1944--1948},
  year={2006},
  publisher={Wiley Online Library},
  doi = {https://doi.org/10.1002/pssc.200565286}
}

@article{mohmad2014localization,
  title={{Localisation effects and band gap of GaAsBi alloys}},
  author={Mohmad, AR and Bastiman, F and Hunter, CJ and Richards, RD and Sweeney, SJ and Ng, JS and David, JPR and Majlis, BY},
  journal={Physica Status Solidi (b)},
  volume={251},
  number={6},
  pages={1276--1281},
  year={2014},
  publisher={Wiley Online Library},
  doi = {https://doi.org/10.1002/pssb.201350311}
}

@book{baranovski2006charge,
  title={Charge Transport in Disordered Solids with Applications in Electronics},
  author={Baranovski, Sergei},
  year={2006},
  publisher={John Wiley \& Sons},
  doi = {10.1002/0470095067}
}

@article{Weisbuch2021,
   author = {Claude Weisbuch and Shuji Nakamura and Yuh Renn Wu and James S. Speck},
   issn = {21928614},
   issue = {1},
   journal = {Nanophotonics},
   month = {1},
   pages = {3-21},
   publisher = {De Gruyter Open Ltd},
   title = {{Disorder effects in nitride semiconductors: Impact on fundamental and device properties}},
   volume = {10},
   year = {2021},
   doi = {https://doi.org/10.1515/nanoph-2020-0590}
}

@article{Masenda2021,
author = {Masenda, Hilary and Schneider, Lorenz Maximilian and Adel Aly, Mohammed and Machchhar, Shachi Jayant and Usman, Arslan and Meerholz, Klaus and Gebhard, Florian and Baranovskii, Sergei D. and Koch, Martin},
title = {Energy Scaling of Compositional Disorder in Ternary Transition-Metal Dichalcogenide Monolayers},
journal = {Advanced Electronic Materials},
volume = {7},
number = {8},
pages = {2100196},
year = {2021},
doi = {https://doi.org/10.1002/aelm.202100196}
}

@article{David-Weisbusch2022,
  title = {{Excitons in a disordered medium: A numerical study in InGaN quantum wells}},
  author = {David, Aurelien and Weisbuch, Claude},
  journal ={Physical Review Research},
  volume = {4},
  issue = {4},
  pages = {043004},
  numpages = {21},
  year = {2022},
  month = {Oct},
  publisher = {American Physical Society},  
  doi = {https://doi.org/10.1103/PhysRevResearch.4.043004}
}

@article{haller2018gan,
  title={{GaN surface as the source of non-radiative defects in InGaN/GaN quantum wells}},
  author={Haller, C and Carlin, J-F and Jacopin, Gwenol{\'e} and Liu, W and Martin, D and Butt{\'e}, R and Grandjean, N},
  journal={Applied Physics Letters},
  volume={113},
  number={11},
  year={2018},
  publisher={AIP Publishing},
  url = {https://doi.org/10.1063/1.5048010}
}

@article{chen2005effects,
  title={{Effects of silicon doping on the nanostructures of InGaN/GaN quantum wells}},
  author={Chen, Meng-Ku and Cheng, Yung-Chen and Chen, Jiun-Yang and Wu, Cheng-Ming and Yang, CC and Ma, Kung-Jen and Yang, Jer-Ren and Rosenauer, Andreas},
  journal={Journal of Crystal Growth},
  volume={279},
  number={1-2},
  pages={55--64},
  year={2005},
  publisher={Elsevier},
  url = {https://doi.org/10.1016/j.jcrysgro.2005.02.018}
}

@article{Rajan2006,
    author = {Rajan, S. and DenBaars, S.P. and Mishra, U.K. and Xing, H. and Jena, D.},
    title = {{Electron mobility in graded AlGaN alloys}},
    journal = {Applied Physics Letters},
    volume = {88},
    number = {4},
    pages = {042103},
    year = {2006},
    month = {01},
    doi = {https://doi.org/10.1063/1.2165190}
}

@article{Brown2010,
    title = {{Finite element simulations of compositionally graded InGaN solar cells}},
    journal = {Solar Energy Materials and Solar Cells},
    volume = {94},
    number = {3},
    pages = {478-483},
    year = {2010},
    author = {G.F. Brown and J.W. Ager and W. Walukiewicz and J. Wu},
    doi = {10.1016/j.solmat.2009.11.010}
}

@article{grandpierron2024understanding,
  title={{Understanding and quantifying the benefit of graded aluminium gallium nitride channel high-electron mobility transistors}},
  author={Grandpierron, Fran{\c{c}}ois and Carneiro, Elodie and Ben-Hammou, Lyes and Moon, Jeong-Sun and Medjdoub, Farid},
  journal={Micromachines},
  doi = {https://doi.org/10.3390/mi15111356},
  volume={15},
  number={11},
  pages={1356},
  year={2024},
  publisher={MDPI}
}

@article{zhang2021compositionally,
  title={{Compositionally graded III-nitride alloys: Building blocks for efficient ultraviolet optoelectronics and power electronics}},
  author={Zhang, H. and Huang, C. and Song, K. and Yu, H. and Xing, C. and Wang, D. and Liu, Z. and Sun, H.},
  journal={Reports on Progress in Physics},
  volume={84},
  number={4},
  pages={044401},
  year={2021},
  publisher={IOP Publishing},
  doi = {10.1088/1361-6633/abde93}
}

@article{zhao2023toward,
  title={{Toward high efficiency at high temperatures: Recent progress and prospects on InGaN-based solar cells}},
  author={Zhao, Y. and Xu, M. and Huang, X. and Lebeau, J. and Li, T. and Wang, D. and Fu, H. and Fu, K. and Wang, X. and Lin, J. and others},
  journal={Materials Today Energy},
  volume={31},
  pages={101229},
  year={2023},
  publisher={Elsevier},
  doi = {https://doi.org/10.1016/j.mtener.2022.101229}
}

@article{Prabakaran2022,
title = {{Effect of spiral-like islands on structural quality, optical and electrical performance of InGaN/GaN heterostructures grown by metal organic chemical vapour deposition}},
journal = {Materials Science in Semiconductor Processing},
volume = {142},
pages = {106479},
year = {2022},
author = {K. Prabakaran and R. Ramesh and P. Arivazhagan and M. Jayasakthi and S. Sanjay and S. Surender and I. {Davis Jacob} and M. Balaji and K. Baskar},
doi = {https://doi.org/10.1016/j.mssp.2022.106479}
}

@article{Adhikari2016,
    author = {Adhikari, R. and Li, T. and Capuzzo, G. and Bonanni, A.},
    title = {{Controlling a three dimensional electron slab of graded Al$_{x}$Ga$_{1-x}$N}},
    journal = {Applied Physics Letters},
    volume = {108},
    number = {2},
    pages = {022105},
    year = {2016},
    month = {01},
    doi = {https://doi.org/10.1063/1.4939788}
}

@article{Sang2021,
    author = {Sang, L. and Sumiya, M. and Liao, M. and Koide, Y. and Yang, X. and Shen, B.},
    title = {{Polarisation-induced hole doping for long-wavelength In-rich InGaN solar cells}},
    journal = {Applied Physics Letters},
    volume = {119},
    number = {20},
    pages = {202103},
    year = {2021},
    month = {11},
    doi = {https://doi.org/10.1063/5.0071506}
}

@article{Amano1989,
   author = {Hiroshi Amano and Masahiro Kito and Kazumasa Hiramatsu and Isamu Akasaki},
    issn = {13474065},
   issue = {12 A},
   journal = {Japanese Journal of Applied Physics},
   month = {12},
   pages = {L2112-L2114},
   publisher = {IOP Publishing},
   title = {{P-type conduction in Mg-doped GaN treated with low-energy electron beam irradiation (LEEBI)}},
   volume = {28},
   year = {1989},
   doi = {10.1143/JJAP.28.L2112}
}

@article{Nakamura1992Hole,
   author = {Shuji Nakamura and Naruhito Iwasa and Masayuki Senoh and Takashi Mukai},
   issn = {13474065},
   issue = {5 R},
   journal = {Japanese Journal of Applied Physics},
   month = {5},
   pages = {1258-1266},
   publisher = {IOP Publishing},
   title = {{Hole compensation mechanism of p-type GaN films}},
   volume = {31},
   year = {1992},
   doi = {10.1143/JJAP.31.1258}
}

@article{dietl2000zener,
  title= {{Zener model description of ferromagnetism in zinc-blende magnetic semiconductors}},
  author={Dietl, T. and Ohno, H. and Matsukura, F. and Cibert, J. and Ferrand, D.},
  journal={Science},
  volume={287},
  number={5455},
  pages={1019--1022},
  year={2000},
  publisher={American Association for the Advancement of Science},
  doi = {https://doi.org/10.1126/science.287.5455.1019}
}

@Inbook{Bonanni2021,
    author={Bonanni, Alberta and Dietl, Tomasz and Ohno, Hideo},
    editor={Coey, J. M. D. and Parkin, Stuart S.P.},
    title={{Dilute Magnetic Materials}},
    bookTitle={Handbook of Magnetism and Magnetic Materials},
    year={2021},
    publisher={Springer International Publishing},
    address={Cham},
    pages={923--978},
    doi = {https://doi.org/10.1007/978-3-030-63210-6_21}
}

@article{Dietl2014,
   author = {T. Dietl and H. Ohno},
   issue = {1},
   journal = {Reviews of Modern Physics},
   pages = {187-251},
   title = {{Dilute ferromagnetic semiconductors: Physics and spintronic structures}},
   volume = {86},
   year = {2014},
   doi = {https://doi.org/10.1103/RevModPhys.86.187}
}

@article{Dietl2019,
   author = {T. Dietl and A. Bonanni and H. Ohno},
   issue = {8},
   journal = {Journal of Semiconductors},
   month = {8},
   pages = {080301},
   publisher = {Institute of Physics Publishing},
   title = {{Families of magnetic semiconductors - An overview}},
   volume = {40},
   year = {2019},
   doi = {10.1088/1674-4926/40/8/080301}
}

@article{moon1999optical,
  title={{Optical and structural studies of phase separation in InGaN film grown by MOCVD}},
  author={Moon, Yong-Tae and Kim, Dong-Joon and Song, Keun-Man and Lee, In-Hwan and Yi, Min-Su and Noh, Do-Young and Choi, Chel-Jong and Seong, Tae-Yeon and Park, Seong-Ju},
  journal={Physica Status Solidi (b)},
  volume={216},
  number={1},
  pages={167--170},
  year={1999},
  publisher={Wiley Online Library},
  doi = {https://doi.org/10.1002/(SICI)1521-3951(199911)216:1%3C167::AID-PSSB167%3E3.0.CO;2-G}
}

@article{li2009suppression,
  title={{Suppression of phase separation in InGaN layers grown on lattice-matched ZnO substrates}},
  author={Li, Nola and Wang, Shen-Jie and Park, Eun-Hyun and Feng, Zhe Chuan and Tsai, Hung-Lin and Yang, Jer-Ren and Ferguson, Ian},
  journal={Journal of Crystal Growth},
  volume={311},
  number={22},
  pages={4628--4631},
  year={2009},
  publisher={Elsevier},
  doi = {https://doi.org/10.1016/j.jcrysgro.2009.09.004}
}

@article{davies2016,
  author = {Davies, M.J. and Hammersley, S. and Massabuau, F.C.-P. and Dawson, P. and Oliver, R.A. and Kappers, M.J. and Humphreys, C.J.},
  title = {{A comparison of the optical properties of InGaN/GaN multiple quantum well structures grown with and without Si-doped InGaN prelayers}},
  journal = {Journal of Applied Physics},
  volume = {119},
  number = {5},
  year = {2016},
  url = {https://doi.org/10.1063/1.4941321},
}

@article{yamamoto2013,
   author={Yamamoto, A. and Mihara, A. and Shigekawa, N. and Narita, N.},
   title={{Marked suppression of In incorporation in heavily Si-doped In$_x$Ga$_{1-x}$N ($x \sim$ 0.3) grown on GaN/$\alpha$-Al2O3 (0001) template}},
  journal = {Applied Physics Letters},
  volume = {103},
  number = {8},
  year = {2013},
  url = {https://doi.org/10.1063/1.4819075},
}

@article{pandey2013band,
  title={{Band bowing and Si donor levels in InGaN layers investigated by surface photo voltage spectroscopy}},
  author={Pandey, S and Cavalcoli, Daniela and Cavallini, Anna},
  journal={Applied Physics Letters},
  volume={102},
  number={14},
  year={2013},
  publisher={AIP Publishing},
  url = {https://doi.org/10.1063/1.4799658}
}

@article{ryu2000silicon,
  title={{Silicon doping effect on the optical properties of In$_{0.15}$Ga$_{0.85}$N/In$_{0.015}$Ga$_{0.985}$N quantum wells}},
  author={Ryu, M-Y and Yu, YJ and Yu, PW and Lee, JI and Yu, SK and Oh, ES and Nam, OH and Sone, CS and Park, YJ and Kim, TI and others},
  journal={Solid State Communications},
  volume={116},
  number={12},
  pages={675--678},
  year={2000},
  publisher={Elsevier},
  doi = {https://doi.org/10.1016/S0038-1098(00)00410-5}
}

@article{Varshni1967,
   author = {Y. P. Varshni},
   issn = {00318914},
   issue = {1},
   journal = {Physica},
   month = {1},
   pages = {149-154},
   publisher = {North-Holland},
   title = {Temperature dependence of the energy gap in semiconductors},
   volume = {34},
   year = {1967},
  doi = {https://doi.org/10.1016/0031-8914(67)90062-6}
}

@article{wang2006band,
  title={{Band gap renormalisation and carrier localisation effects in InGaN/ GaN quantum-wells light emitting diodes with Si doped barriers}},
  author={Wang, YJ and Xu, SJ and Li, Q and Zhao, DG and Yang, H},
  journal={Applied Physics Letters},
  volume={88},
  number={4},
  year={2006},
  publisher={AIP Publishing},
  url = {https://doi.org/10.1063/1.2168035}
}

@article{cho1998influence,
  title={{Influence of Si doping on characteristics of InGaN/GaN multiple quantum wells}},
  author={Cho, Yong-Hoon and Song, JJ and Keller, S and Minsky, MS and Hu, E and Mishra, UK and DenBaars, SP},
  journal={Applied Physics Letters},
  volume={73},
  number={8},
  pages={1128--1130},
  year={1998},
  publisher={American Institute of Physics},
  doi = {https://doi.org/10.1063/1.122105}
}

@article{cheng2003quantum,
  title={{Quantum dot formation in InGaN/GaN quantum well structures with silicon doping and the mechanisms for radiative efficiency improvement}},
  author={Cheng, Yung-Chen and Feng, Shih-Wei and Lin, En-Chiang and Yang, Chih-Chung and Tseng, Cheng-Hua and Hsu, Cheng and Ma, Kung-Jen},
  journal={Physica Status Solidi (c)},
  number={4},
  pages={1093--1096},
  year={2003},
  publisher={Wiley Online Library},
  doi ={https://doi.org/10.1002/pssc.200303001}
}

@article{cho1999influence,
  title={{Influence of Si-doping on carrier localisation of MOCVD-grown InGaN/GaN multiple quantum wells}},
  author={Cho, Yong-Hoon and Schmidt, TJ and Bidnyk, S and Song, JJ and Keller, S and Mishra, UK and DenBaars, SP},
  journal={MRS Internet Journal of Nitride Semiconductor Research},
  volume={4},
  number={Suppl 1},
  pages={715--720},
  year={1999},
  publisher={Springer},
  doi = {https://doi.org/10.1557/S1092578300003306}
}

@article{li2017localization,
  title={{Localisation landscape theory of disorder in semiconductors. III. Application to carrier transport and recombination in light-emitting diodes}},
  author={Li, Chi-Kang and Piccardo, Marco and Lu, Li-Shuo and Mayboroda, Svitlana and Martinelli, Lucio and Peretti, Jacques and Speck, James S and Weisbuch, Claude and Filoche, Marcel and Wu, Yuh-Renn},
  journal={Physical Review B},
  volume={95},
  number={14},
  pages={144206},
  year={2017},
  publisher={APS},
  doi = {https://doi.org/10.1103/PhysRevB.95.144206}
}

@article{piccardo2017localization,
  title={{Localisation landscape theory of disorder in semiconductors. II. Urbach tails of disordered quantum well layers}},
  author={Piccardo, Marco and Li, Chi-Kang and Wu, Yuh-Renn and Speck, James S and Bonef, Bastien and Farrell, Robert M and Filoche, Marcel and Martinelli, Lucio and Peretti, Jacques and Weisbuch, Claude},
  journal={Physical Review B},
  volume={95},
  number={14},
  pages={144205},
  year={2017},
  publisher={APS},
  doi = {https://doi.org/10.1103/PhysRevB.95.144205}
}

@article{filoche2017localization,
  title={{Localisation landscape theory of disorder in semiconductors. I. Theory and modeling}},
  author={Filoche, Marcel and Piccardo, Marco and Wu, Yuh-Renn and Li, Chi-Kang and Weisbuch, Claude and Mayboroda, Svitlana},
  journal={Physical Review B},
  volume={95},
  number={14},
  pages={144204},
  year={2017},
  publisher={APS},
  doi = {https://doi.org/10.1103/PhysRevB.95.144204}
}

@article{akasaki2015nobel,
  title={Nobel Lecture: Fascinated journeys into blue light},
  author={Akasaki, Isamu},
  journal={Reviews of Modern Physics},
  volume={87},
  number={4},
  pages={1119--1131},
  year={2015},
  publisher={APS},
  doi = { https://doi.org/10.1103/RevModPhys.87.1119}
}

@article{amano2015nobel,
  title={{Nobel Lecture: Growth of GaN on sapphire via low-temperature deposited buffer layer and realisation of p-type GaN by Mg doping followed by low-energy electron beam irradiation}},
  author={Amano, Hiroshi},
  journal={Reviews of Modern Physics},
  volume={87},
  number={4},
  pages={1133--1138},
  year={2015},
  publisher={APS},
  doi = {https://doi.org/10.1103/RevModPhys.87.1133}
}

@article{gebhard2023quantum,
  title={{Quantum states in disordered media. I. Low-pass filter approach}},
  author={Gebhard, F and Nenashev, AV and Meerholz, K and Baranovskii, SD},
  journal={Physical Review B},
  volume={107},
  number={6},
  pages={064206},
  year={2023},
  publisher={APS},
  doi = {https://doi.org/10.1103/PhysRevB.107.064206}
}

@article{nenashev2023quantum,
  title={{Quantum states in disordered media. II. Spatial charge carrier distribution}},
  author={Nenashev, AV and Baranovskii, SD and Meerholz, K and Gebhard, F},
  journal={Physical Review B},
  volume={107},
  number={6},
  pages={064207},
  year={2023},
  publisher={APS},
  doi = {https://doi.org/10.1103/PhysRevB.107.064207}
}

@article{nakamura1998,
    author = {Shuji Nakamura},
    title = {{The roles of structural imperfections in InGaN-based blue light-emitting diodes and laser diodes}},
    journal = {Science},
    volume = {281},
    number = {5379},
    pages = {956-961},
    year = {1998},
    doi = {10.1126/science.281.5379.956},
}

@article{wu2021Anderson,
    author = {Wu, Meng-Jer and Wu, Shang-Cheng and Shen, Tien-Lin and Liao, Yu-Ming and Chen, Yang-Fang},
    title = {{Anderson localisation enabled spectrally stable deep-ultraviolet laser based on metallic nanoparticle decorated AlGaN multiple quantum wells}},
    journal = {ACS Nano},
    volume = {15},
    number = {1},
    pages = {330-337},
    year = {2021},
    doi = {10.1021/acsnano.0c04512},
}

@article{Bhunia2024,
    author = {Bhunia, Swagata and Majumder, Ayan and Chatterjee, Soumyadip and Sarkar, Ritam and Nag, Dhiman and Saha, Kasturi and Mahapatra, Suddhasatta and Laha, Apurba},
    title = {{Ultrafast green single photon emission from an InGaN quantum dot-in-a-GaN nanowire at room temperature}},
    journal = {Applied Physics Letters},
    volume = {125},
    number = {4},
    pages = {044002},
    year = {2024},
    month = {07},
    doi = {10.1063/5.0213596},
}

@article{Xue2025,
    author = {Xue, Zhaoyi and Su, Zhicheng and Xu, Shijie},
    title = {A modified model for thermodynamics of localized-state luminescence in disordered materials},
    journal = {Journal of Applied Physics},
    volume = {138},
    number = {4},
    pages = {045705},
    year = {2025},
    month = {07},
    doi = {10.1063/5.0276875},
    url = {https://doi.org/10.1063/5.0276875},
 }

@article{Moss1954,
    doi = {10.1088/0370-1301/67/10/306},
    url = {https://doi.org/10.1088/0370-1301/67/10/306},
    year = {1954},
    month = {oct},
    publisher = {},
    volume = {67},
    number = {10},
    pages = {775},
    author = {T S Moss},
    title = {The Interpretation of the Properties of Indium Antimonide},
    journal = {Proceedings of the Physical Society. Section B},
}

@article{Burstein1954,
    title = {{Anomalous optical absorption limit in InSb}},
    author = {Burstein, Elias},
    journal = {Physical Review},
    volume = {93},
    issue = {3},
    pages = {632--633},
    numpages = {0},
    year = {1954},
    month = {Feb},
    publisher = {American Physical Society},
    doi = {10.1103/PhysRev.93.632},
    url = {https://link.aps.org/doi/10.1103/PhysRev.93.632}
}

@article{Lee2010,
    author = {Lee, Sung-Nam and Kim, Jihoon and Kim, Kyoung-Kook and Kim, Hyunsoo and Kim, Han-Ki},
    title = {{Thermal stability of Si-doped InGaN multiple-quantum wells for high efficiency light emitting diodes}},
    journal = {Journal of Applied Physics},
    volume = {108},
    number = {10},
    pages = {102813},
    year = {2010},
    month = {11},
    url = {https://doi.org/10.1063/1.3511712},
}

@article{delaney2009auger,
  title={{Auger recombination rates in nitrides from first principles}},
  author={Delaney, Kris T and Rinke, Patrick and Van de Walle, Chris G},
  journal={Applied Physics Letters},
  volume={94},
  number={19},
  year={2009},
  publisher={AIP Publishing},
  url = {https://doi.org/10.1063/1.3133359}
}

@article{thaalbi2025streamlined,
  title={{Streamlined MOCVD growth of red InGaN LEDs via precursor-mediated surface reconstruction}},
  author={Thaalbi, Hamza and Kim, Baul and Park, Sohyeon and Abdullah, Ameer and Kulkarni, Mandar A and Tariq, Fawad and Din, Haseeb Ud and Jun, Seongmoon and Jang, Ho Won and Cho, Yong-Hoon and others},
  journal={Surfaces and Interfaces},
  volume={65},
  pages={106488},
  year={2025},
  publisher={Elsevier},
  doi = {https://doi.org/10.1016/j.surfin.2025.106488}
}

@article{cao2019carrier,
  title={{Carrier dynamics determined by carrier-phonon coupling in InGaN/GaN multiple quantum well blue light emitting diodes}},
  author={Cao, Sheng and Wu, Xiao-Ming and Liu, Jun-Lin and Jiang, Feng-Yi},
  journal={Chinese Physics Letters},
  volume={36},
  number={2},
  pages={028501},
  year={2019},
  publisher={IOP Publishing},
  doi = {10.1088/0256-307X/36/2/028501}
}

@article{callsen2019probing,
  title={{Probing alloy formation using different excitonic species: the particular case of InGaN}},
  author={Callsen, Gordon and Butt{\'e}, Raphael and Grandjean, Nicolas},
  journal={Physical Review X},
  volume={9},
  number={3},
  pages={031030},
  year={2019},
  publisher={APS},
  doi = {https://doi.org/10.1103/PhysRevX.9.031030}
}

@article{xu2002direct,
  title={{Direct determination of free exciton binding energy from phonon-assisted luminescence spectra in GaN epilayers}},
  author={Xu, SJ and Liu, W and Li, MF},
  journal={Applied Physics Letters},
  volume={81},
  number={16},
  pages={2959--2961},
  year={2002},
  publisher={American Institute of Physics},
  doi = {https://doi.org/10.1063/1.1514391}
}

@article{park2024ingan,
  title={{InGaN-based blue and red micro-LEDs: Impact of carrier localization}},
  author={Park, Jeong-Hwan and Pristovsek, Markus and Han, Dong-Pyo and Kim, Bumjoon and Lee, Soo Min and Hanser, Drew and Parikh, Pritesh and Cai, Wentao and Shim, Jong-In and Lee, Dong-Seon and others},
  journal={Applied Physics Reviews},
  volume={11},
  number={4},
  year={2024},
  publisher={AIP Publishing},
  url = {https://doi.org/10.1063/5.0195261}
}

@article{kang2025growth,
  title={{Growth of compositionally uniform In$_x$Ga$_{1-x}$N layers with low relaxation degree on GaN by molecular beam epitaxy}},
  author={Kang, Jingxuan and Ruiz, Mikel G{\'o}mez and Van Dinh, Duc and Campbell, Aidan F and John, Philipp and Auzelle, Thomas and Trampert, Achim and L{\"a}hnemann, Jonas and Brandt, Oliver and Geelhaar, Lutz},
  journal={Journal of Physics D: Applied Physics},
  volume={58},
  number={14},
  pages={14LT01},
  year={2025},
  publisher={IOP Publishing},
  doi = {10.1088/1361-6463/adb4e7}
}

@article{dang2018threading,
  title={{Threading dislocation density effect on the electrical and optical properties of InGaN light-emitting diodes}},
  author={Dang, Suihu and Li, Chunxia and Lu, Mengchun and Guo, Hongli and He, Zelong},
  journal={Optik},
  volume={155},
  pages={26--30},
  year={2018},
  publisher={Elsevier},
  doi = {https://doi.org/10.1016/j.ijleo.2017.10.096}
}

@article{xie2017inhomogeneous,
  title={Inhomogeneous composition distribution in monolayer transition metal dichalcogenide alloys},
  author={Xie, Shuang and Xu, Mingsheng and Huang, Shuyun and Liang, Tao and Wang, Shengping and Li, Hongfei and Iwai, Hideo and Onishi, Keiko and Hanagata, Nobutaka and Fujita, Daisuke and others},
  journal={Materials Research Express},
  volume={4},
  number={4},
  pages={045004},
  year={2017},
  publisher={IOP Publishing},
   doi = {https://doi.org/10.1088/2053-1591/aa6859}
}

@article{ozturk2014microstructural,
  title={{Microstructural properties of InGaN/GaN light-emitting diode structures with different In content grown by MOCVD}},
  author={{\"O}zt{\"u}rk, MK and {\c{C}}{\"o}rek{\c{c}}i, S and Tamer, M and {\c{C}}etin, S{\c{S}} and {\"O}z{\c{c}}elik, S and {\"O}zbay, E},
  journal={Applied Physics A},
  volume={114},
  number={4},
  pages={1215--1221},
  year={2014},
  publisher={Springer},
  doi ={10.1007/s00339-013-7857-2}
}

@article{harrington2021back,
  title={Back-to-Basics tutorial: X-ray diffraction of thin films},
  author={Harrington, George F and Santiso, Jos{\'e}},
  journal={Journal of Electroceramics},
  volume={47},
  number={4},
  pages={141--163},
  year={2021},
  publisher={Springer},
  doi = {https://doi.org/10.1007/s10832-021-00263-6}
}

@article{kim2010analysis,
  title={{Analysis of time-resolved photoluminescence of InGaN quantum wells using the carrier rate equation}},
  author={Kim, Hyunsung and Shin, Dong-Soo and Ryu, Han-Youl and Shim, Jong-In},
  journal={Japanese Journal of Applied Physics},
  volume={49},
  number={11R},
  pages={112402},
  year={2010},
  publisher={IOP Publishing},
  doi = {10.1143/JJAP.49.112402}
}

@article{yu2022invalidation,
  title={{Invalidation of the acquisition of internal quantum efficiency using temperature-dependent photoluminescence in InGaN quantum wells with high threading dislocation density}},
  author={Yu, Jiadong and Wang, Lai and Wang, Jian and Hao, Zhibiao and Luo, Yi and Sun, Changzheng and Han, Yanjun and Xiong, Bing and Li, Hongtao},
  journal={Journal of Physics D: Applied Physics},
  volume={55},
  number={19},
  pages={195107},
  year={2022},
  publisher={IOP Publishing},
  doi = {10.1088/1361-6463/ac5149}
}

@inbook{Roccaforte2020,
author = {Roccaforte, Fabrizio and Leszczynski, Mike},
publisher = {John Wiley \& Sons, Ltd},
title = {{Introduction to Gallium Nitride Properties and Applications}},
booktitle = {Nitride Semiconductor Technology},
chapter = {1},
pages = {1-39},
doi = {https://doi.org/10.1002/9783527825264.ch1},
year = {2020},
}

@article{Moram2009,
doi = {https://doi.org/10.1088/0034-4885/72/3/036502},
url = {https://doi.org/10.1088/0034-4885/72/3/036502},
year = {2009},
month = {feb},
publisher = {},
volume = {72},
number = {3},
pages = {036502},
author = {Moram, M A and Vickers, M E},
title = {{X-ray diffraction of {III}-nitrides}},
journal = {Reports on Progress in Physics},
}

@article{Ourmazd1984,
author = {A. Ourmazd},
title = {{The electrical properties of dislocations in semiconductors}},
journal = {Contemporary Physics},
volume = {25},
number = {3},
pages = {251--268},
year = {1984},
publisher = {Taylor \& Francis},
doi = {https://doi.org/10.1080/00107518408210707},
URL = {https://doi.org/10.1080/00107518408210707}
}

@incollection{You2009,
author = {J.H. You and H.T. Johnson},
title = {{Chapter 3: Effect of dislocations on electrical and optical properties in GaAs and GaN}},
booktitle = {Solid State Physics},
publisher = {Academic Press},
volume = {61},
pages = {143-261},
year = {2009},
issn = {0081-1947},
doi = {https://doi.org/10.1016/S0081-1947(09)00003-4}
}

@article{Albrecht2014,
  title = {{Origin of the unusually strong luminescence of $a$-type screw dislocations in GaN}},
  author = {Albrecht, M. and Lymperakis, L. and Neugebauer, J.},
  journal = {Physical Review B},
  volume = {90},
  issue = {24},
  pages = {241201},
  numpages = {4},
  year = {2014},
  month = {Dec},
  publisher = {American Physical Society},
  doi = {https://doi.org/10.1103/PhysRevB.90.241201},
  url = {https://link.aps.org/doi/10.1103/PhysRevB.90.241201}
}

@article{Mishra2011,
author = {Mishra, K. C. and Johnson, K. H. and Schmidt, P. C.},
title = {{First-principles investigation of the electronic structures of edge dislocations in GaN}},
journal = {Physica Status Solidi (a)},
volume = {208},
number = {7},
pages = {1555-1557},
doi = {https://doi.org/10.1002/pssa.201000925},
url = {https://onlinelibrary.wiley.com/doi/abs/10.1002/pssa.201000925},
year = {2011}
}

@article{Bharadwaj2019,
    author = {Bharadwaj, Shyam and Islam, S. M. and Nomoto, Kazuki and Protasenko, Vladimir and Chaney, Alexander and Xing, Huili (Grace) and Jena, Debdeep},
    title = {Bandgap narrowing and {Mott} transition in {Si}-doped {Al$_{0.7}$Ga$_{0.3}$N}},
    journal = {Applied Physics Letters},
    volume = {114},
    number = {11},
    pages = {113501},
    year = {2019},
    month = {03},
    issn = {0003-6951},
    doi = {10.1063/1.5086052},
    url = {https://doi.org/10.1063/1.5086052},
}

@book{Mott1990,
  author    = {Mott, Nevill F.},
  title     = {Metal-insulator transitions},
  edition   = {Second},
  publisher = {CRC Press},
  address   = {London},
  year      = {1990},
  isbn      = {978-0850667837},
  doi       = {10.1201/b12795},
  url       = {https://doi.org/10.1201/b12795}
}

@article{Nishikawa2023,
    author = {Nishikawa, Yuto and Ueno, Kohei and Kobayashi, Atsushi and Fujioka, Hiroshi},
    title = {Preparation of degenerate n-type {Al$_{x}$Ga$_{1−x}$N} ($0 < x  \leq 0.81$) with record low resistivity by pulsed sputtering deposition},
    journal = {Applied Physics Letters},
    volume = {122},
    number = {23},
    pages = {232102},
    year = {2023},
    month = {06},
    issn = {0003-6951},
    doi = {10.1063/5.0144418},
    url = {https://doi.org/10.1063/5.0144418},
}

@incollection{shim2013active,
  title={Active region part b. internal quantum efficiency},
  author={Shim, Jong-In},
  booktitle={III-Nitride Based Light Emitting Diodes and Applications},
  pages={153--195},
  year={2013},
  publisher={Springer},
  doi = {https://doi.org/10.1007/978-94-007-5863-6_7}
}
\end{document}